\documentclass[twocolumn,showpacs,preprintnumbers,amsmath,amssymb, superscriptaddress]{revtex4-1}

\usepackage{graphicx}
\usepackage{bm}

\begin{document}

\title{The critical 2-dimensional Ising model with fixed boundaries}

\author{Xintian Wu}
\email{wuxt@bnu.edu.cn}
\affiliation{Department of Physics, Beijing Normal University,
Beijing, 100875, China}

\author{Nickolay Izmailyan}
\email{izmail@yerphi.am}
\affiliation{Applied Mathematics Research Centre, Coventry University, Coventry CV1
5FB, United Kingdom} \affiliation{A.I. Alikhanyan National
Science Laboratory, Alikhanian Br.2, 375036 Yerevan, Armenia.  }

\date{\today}

\begin{abstract}
The critical 2-dimensional Ising model is studied with four types boundary conditions: free, fixed ferromagnetic, fixed antiferromagnetic, fixed double antiferromagnetic.  Using Bond Propagation algorithms with surface fields, we obtained the free energy , internal energy and specific heat numerically on square lattices with square shape and various combinations of the four types boundary conditions. The numerical data are analyzed with finite size scaling. The bulk, edge and corner terms are extracted very accurately. The exact results are conjectured for the corner logarithmic term in the free energy, the edge and corner logarithmic terms in the internal energy and specific heat. The corner logarithmic terms in the free energy agree with the conformal field theory very well.

\end{abstract}


\maketitle

\section{Introduction}

Two dimensional Ising model should be the best understood statistical model. Exact results of the model on finite sizes with various boundary conditions (BCs) have been studied \cite{onsager,kaufman,newell,fisher1969,fisher,brascamp,izmailian2002a,izmailian2002b,izmailian2007,salas,Janke}.
Detailed knowledge has been obtained for the torus case \cite{izmailian2002a,salas}, for helical BCs
\cite{izmailian2007}, for Brascamp-Kunz BCs \cite{brascamp,izmailian2002b,Janke} and for infinitely long cylinder\cite{izmailian1}. The Ising model on the finite lattice plays an important role on the finite size scaling \cite{privman,privman1}, which finds extensive applications in the analysis of experimental, Monte Carlo, and transfer-matrix data, as well as in recent theoretical developments related to conformal invariance \cite{blote,cardy,kleban}. However for the complete open BCs, i.e. with open edges and sharp corners, some interesting results begin to appear in recent years \cite{jacobsen, wu, wu2}. In these works, the free BCs are studied. In this paper, we consider the fixed BCs.

One of the methods to solve the two dimensional Ising model with free boundary is the bond propagation (BP) algorithm. It was developed for computing the partition function of the Ising model in two dimensions first \cite{loh1,loh2}. Later on, the BP algorithm for the internal energy and specific heat are also developed \cite{wu3}. It is so powerful that very large system sizes and very accurate numerical results can be reached. With this algorithm, the calculations have been carried out on square and triangular lattices with free boundaries \cite{wu,wu2,wu3}. The results of free energy, internal energy and specific heat are surprisingly accurate to $10^{-26}$. The exact edge and corner terms on the square lattice and triangle lattice with various shapes are conjectured.

Recently one of the authors develops the BP algorithm with a surface field \cite{wu4}, which is called SFBP algorithm. We use the SFBP algorithm to study the two dimensional Ising model with fixed boundary. We have studied three types of fixed BCs: ``$\pm$", ``$a$" and ``$b$", which are defined in section II. The free BC is denoted by ``$0$". We have carried out numerical calculations of the free energy, internal energy and specific heat on the square lattice with a square shape. For each edge, we assign a type of BC. We present the results of ten BCs including $(++++)$  (where the four edges are assigned ``$+$" type BC), $(aaaa)$, $(bbbb)$, $(+-+-)$,$(+0+0)$,$(+a+a)$, $(+b+b)$, $(a0a0)$, $(b0b0)$, and $(abab)$.   Through fitting the data, we get very accurate expansions of the free energy, internal energy and specific heat. The corner logarithmic corrections in the free energy verify the conformal field theory (CFT) predictions \cite{imamura,bondesan,roberto}. The exact edge and corner logarithmic terms in the internal energy and specific heat are conjectured.

The accurate finite size expansions at the critical point are helpful to the CFT and renormalization group (RG) study on the boundary effects \cite{vicari}.
An interesting way to  see the effect of the irrelevant operators is to compute the asymptotic expansion (in powers of $L^{-1}$) of the free energy and its derivatives with respect to the temperature at the critical point. In the corner terms of the free energy and in the edge and corner terms of the internal energy and specific heat, there are logarithmic corrections. The logarithmic terms are usually related to the singular part of free energy \cite{privman}. Therefore studying these logarithmic corrections is specially important.

Our paper is organized as follows. Since this paper is very long, we first summarize our results in section II.  In section III, IV and V,  we present the fittings of the free energy density,  the internal energy and the specific heat respectively. Section VI is a discussion and Acknowledgement.

\section{The critical 2-dimensional Ising model with mixed BCS}

The Fig. 1 shows some lattices with typical BCs. We have studied three types of fixed BCs. The first one is ``$+$" (``$-$") type, where all the spins at the boundary are fixed to be positive (negative). The second one is antiferromagnetic ``a" type, where the spins at the boundary are fixed positive, negative alternatively. The third one is double antiferromagnetic "b", where the spins at the boundary are fixed positive, positive, negative, negative successively. Together with the free boundary (denoted by ``0"), we have four types BCs. For a square, there are four edges, we denote the boundaries by ($\alpha_1,\alpha_2,\alpha_3,\alpha_4$), where $\alpha_i=0,+,-,a,b$. For example, as shown in Fig. (1a) the BCs on the four edges are fixed ``+", fixed ``a", fixed ``b", fixed ``-" in proper order respectively, then we denote the BCs by ``$(+ab-)$".

We study the Ising model with these BCs using SFBP algorithms \cite{wu4}. The fixed BCs is realized in SFBP in the following way. Consider a $(N+2)\times (N+2)$ lattice with coupling constant between spins being $J$. If the bottom  (the zeroth row of spins) is assigned to be ``$+$" type boundary, it is equivalent to applying a field with intensity of $J$ to the first row of spins without the zeroth row of spins.  If the boundary is the type ``$0$", the surface field on the first row are zero. If the boundary is type ``$a$", the surface field on the first row of spins are $+J$ and $-J$ alternatively. If the boundary is type ``$b$", the surface field on the first row of spins are $+J$, $+J$ $-J$, and $-J$ successively. The other edges can be dealt similarly. Then the $(N+2)\times (N+2)$ lattice with fixed BCs becomes a $N\times N$ lattice with a surface field. In other words we can solve the Ising model with a surface filed to study the fixed boundary. Consider the Ising model with a surface field
\begin{equation}
\mathcal{H}=-J\sum_{<i,j>}\sigma_i\sigma_j-\sum_{i\in \Gamma}H_i\sigma_i
\end{equation}
where $\Gamma$ is the boundary, and the surface field is assigned according to the BCs, which is discussed above. We calculate the partition function
\begin{equation}
Z=\sum_{\{\sigma_i\}}e^{-\beta \mathcal{H}},
\end{equation}
then get the free energy density
\begin{equation}
f=-\frac{\ln Z}{S}
\end{equation}
where $S=N^2$ is the number of spins. The internal energy density is defined by
\begin{equation}
u=-\frac{1}{S}\frac{\partial \ln Z}{\partial \beta}
\label{eq:internal}
\end{equation}
and specific heat per spin by
\begin{equation}
c=\frac{1}{S}\frac{\partial ^2 \ln Z}{\partial \beta ^2}.
\label{eq:speci}
\end{equation}

\begin{figure}

\includegraphics[width=0.5\textwidth]{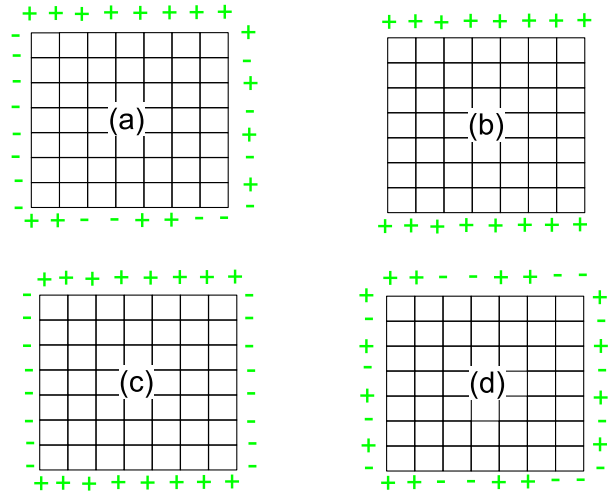}

\caption{ (a) The $8\times 8$ lattice with BCs $(+ab-)$. (b) The $8\times 8$ lattice with BCs $(+0+0)$.
(c) The $8\times 8$ lattice with BCs $(+-+-)$. (d) The $8\times 8$ lattice with BCs $(baba)$. }
\label{boundary}
\end{figure}

We have calculated the free energy, internal energy and specific heat at the critical point $\beta J=\beta_c J=\frac{1}{2}\ln(1+\sqrt{2})$ with various BCs. We only present the results with ten BCs including $(++++),(aaaa),(bbbb)$, $(+a+a),(+-+-),(+b+b),(+0+0)$, $(abab),(0a0a),(0b0b)$.
In the first three cases the BCs of the four edges are the same, so we can get the edge term easily.
In all these cases,  the BCs for the four corners are the same, so we can get the corner term easily.

We find that the critical free energy density can be expanded into
\begin{equation}
f=f_0+\frac{f_s}{N} +
\frac{f_c\ln N }{N^2} +\sum_{k=2}^{k_{max}}\frac{f_k}{N^k}
\label{eq:free}
\end{equation}
where the surface term $f_s=\sum_{i=1}^{4}f_{\rm surf}(\alpha_i)$ is the sum of the four edges' contribution, $\alpha_i=+,-,0,a,b$ denote the BCs of the $i$th edge; the corner term $f_c=\sum_{i=1}^4f_{\rm corn}(\alpha_i\beta_i)$ is the sum of the four corners' contribution, $\alpha_i\beta_i=+-,+0,+a,+b,etc. $ denotes the BCs of the $i$th corner. For example, the contribution from a corner with two edges under BCs ``+" and``0" is denoted by $f_{corn}(+0)$. We expand the the free energy (the internal energy and specific heat in the next two sections) to as high order as possible to make the deviation as small as possible. In all fittings of the free energy, we take $k_{max}=15$.

In our fittings we confirm the exact result given by
Onsager\cite{onsager}, i.e.,
\begin{equation}
f_0 = -\ln  \sqrt{2}-\frac{2}{\pi}G
\label{f-surf}
\end{equation}
where $G=1-\frac{1}{3^2}+\frac{1}{5^2}-\frac{1}{7^2}+\cdots$.

For the free BCs, the edge term $f_{surf}(0)$  has been obtained in our previous work \cite{wu}. It is conjectured that
\begin{equation}
f_{surf}(0)= \frac{1}{2}[\frac{1}{2}\ln(1+\sqrt{2})-D_1]
\end{equation} where
$D_1=\int_{0}^{\pi}\ln[1+\sqrt{2}(1-\cos \theta)^{1/2}(3-\cos\theta)^{-1/2}]$. This is obtained by comparing our numerical result on the rectangle with the exact result on the cylinder by Helen and Fisher \cite{fisher}.

In the fittings for the free energy data,  the other surface terms are given by
\begin{equation}
\begin{array}{lc}
\alpha  &  f_{surf}(\alpha) \\
\pm & -0.200695195538609403401008(3) \\
a & 0.0582601809867433557406(6) \\
b & 0.03597349712225534654(7)
\end{array}
\end{equation}
For $f_{surf}(\pm)$ and $f_{surf}(a)$, it is satisfied that
\begin{equation}
f_{surf}(\pm)+f_{surf}(a) = \frac{1}{2}\ln (1+\sqrt{2})-\frac{2}{\pi}G.
\end{equation}
This is obtained by comparing the result on Brascamp-Kunz BCs \cite{izmailian2002b}, where the cylinder with upper edge under ``$+$" BCs and bottom edge under ``$a$" BCs.

The corner terms $f_{corn}$ are very interesting because they are universal and logarithmic. Cardy and Peschel studied the free energy with in CFT  \cite{cardy}. They predicted that a corner with angle $\pi /2$ and  two edges under free BCs,  gives rise to the term $-\frac{c }{16 }\frac{\ln N}{N^2}$ in the free energy expansion, where $c=1/2$ is the central charge. That is to say $ f_{corn}(00)=-\frac{1}{32}$. This is verified in our previous work on the square lattice and triangle lattice with free BCs \cite{wu,wu2}. Late on, Imamura et al. study the corner terms with different BCs within CFT \cite{imamura,bondesan,roberto}. According to their results, the contribution to the free energy from a corner with two edges under $\alpha$ and $\beta$ BCs, $f_{corn}(\alpha \beta)$ is given by
\begin{equation}
f_{corn}(\alpha\beta) =2\lambda_{\alpha\beta}-\frac{c}{16}
\end{equation}
where $\lambda_{\alpha\beta}$ is the conformal weight of the boundary operator inserting at the corner and $c=1/2$ is the central charge for the Ising model. There are three different conformal weights in the Ising model, namely, $\lambda_{\alpha\beta}=0, 1/16$ and $1/2$. There are also three different conformally invariant boundary conditions, generally denoted by $+$, $-$ and $f$ \cite{Cardy1989}, where first two described fixed boundary conditions on the spin $s=+1$ or $-1$ respectively, and $f$ corresponds to free boundary conditions. When boundary conditions on both sides of the corner are the same ($\alpha=\beta$), the boundary operator is just identity operator with $\lambda=0$. In the case when boundary conditions on one side of the corner is $\alpha=+$ and on another side is $\beta=-$, the boundary operator has conformal weight $\lambda=1/2$ and in the case when boundary condition on one side of the corner is $\alpha=+$ or $-$ and on another side is $\beta=0, a$ or $b$ the boundary operator has conformal weight $\lambda=1/16$.  We also get the following result, when boundary conditions on both sides of the corner are $0, a$ or $b$, the boundary operator has conformal weight $\lambda=0$. Thus we have obtained that the boundary conditions $0, a$ and $b$ are all belong to the free conformally invariant boundary condition. Then we have the table for $f_{corn}(\alpha\beta)$
\begin{equation}
\begin{array}{cccccc}
   & + & - & 0 & a & b \\
+  & -\frac{1}{32} & \frac{31}{32}  &  ~~\frac{3}{32} &  ~~\frac{3}{32}  &   ~~\frac{3}{32}  \\
-  &  & -\frac{1}{32}  &  ~~\frac{3}{32} &  ~~\frac{3}{32}  &   ~~\frac{3}{32}  \\
0  &  &   & -\frac{1}{32}  &  -\frac{1}{32}   &  -\frac{1}{32}   \\
a  &  &   &                &  -\frac{1}{32}  &   -\frac{1}{32}   \\
b  &  &   &                &                 &   -\frac{1}{32}
\end{array}
\label{eq:cft}
\end{equation}
where the first column is $\alpha$ and the first row is $\beta$. It is obvious that $f_{corn}(\alpha \beta)=f_{corn}(\beta \alpha)$.

Our numerical results verified these predictions in very high accuracies. The detailed fittings of the numerical data are presented in section III.
As one can see, the BCs ``$0,a,b$" play the same role in the corner terms. In CFT, these BCs belong to the same type indeed. However they play different roles in the internal energy and specific heat, which are show in the following. This is the reason why we study these three types of BCs ``$0,a,b$".

We find that the critical internal energy  can be expanded into
\begin{equation}
u=u_0+\frac{u_s\ln N}{N} +\frac{u_c\ln N }{N^2} +\sum_{k=1}^{\infty}\frac{u_k}{N^k}
\label{eq:inter}
\end{equation}
where the surface term $u_s=\sum_{i=1}^{4}u_{\rm surf}(\alpha_i)$ is the sum of the four edges' contribution and the corner term $u_c=\sum_{i=1}^4u_{\rm corn}(\alpha_i\beta_i)$ is the sum of the four corners' contribution.

In the expansions, we confirm the exact result of the bulk value \cite{onsager}
\begin{equation}
u_0=-\sqrt{2}
\end{equation}

From the results with various BCs, the surface terms are given by
\begin{eqnarray}
\begin{array}{ccccc}
\alpha & \pm & 0 & a & b   \\
u_{surf}(\alpha) &  -\frac{1}{\pi} &  \frac{1}{\pi} &  \frac{1}{\pi}   & \frac{1}{\pi}
\end{array}
\end{eqnarray}

The corner terms $u_{corn}(\alpha \beta)$ are summarized in the following expression
\begin{equation}
\begin{array}{ccccc}
   & \pm & 0 & a & b \\
\pm  & -\frac{2+\sqrt{2}}{2\pi}  &  \frac{1}{2\pi} &  \frac{\sqrt{2}}{2\pi}  &   \frac{5\sqrt{2}}{8\pi}  \\
0  &                      & \frac{\sqrt{2}}{2\pi}  &  \frac{1}{2\pi}  &    \frac{4-\sqrt{2}}{8\pi}  \\
a  &                      &                        &  \frac{2-\sqrt{2}}{2\pi} &    \frac{8-5\sqrt{2}}{8\pi}  \\
b  &                      &                        &                           &   \frac{4-3\sqrt{2}}{4\pi}
\end{array}
\end{equation}
where $\alpha$ is shown in the left column and $\beta$ is shown in the first row. The corner term $u_{corn}(+-)$ is given by
\begin{equation}
u_{corn}(+-)=-\frac{2+\sqrt{2}}{2\pi}.
\end{equation}
It should be noted that $u_{corn}(+-)=u_{corn}(++)$ as compared with $f_{corn}(+-)\neq f_{corn}(++)$.
The detailed fittings of the numerical data and extracting these coefficients are presented in section IV.

The  critical specific heat can be expanded into
\begin{equation}
c=A_0 \ln N+D_0+\frac{c_s\ln N}{N}
+\frac{c_c\ln N }{N^2} +\sum_{k=1}^{\infty}\frac{c_k}{N^k} \label{eq:specific}
\end{equation}
where the surface term $c_s=\sum_{i=1}^{4}c_{\rm surf}(\alpha_i)$ is the sum of the four edges' contribution, the corner term $c_c=\sum_{i=1}^4c_{\rm corn}(\alpha_i\beta_i)$ is the sum of the four corners' contribution.

In the fittings, we confirm the  Onsager's exact result \cite{onsager}
\begin{equation}
A_0 =\frac{8}{\pi}.
\end{equation}
The surface terms are given by
\begin{eqnarray}
\begin{array}{ccccc}
\alpha & \pm & 0 & a & b   \\
c_{surf}(\alpha) &  \frac{4+\sqrt{2}}{\pi} & \frac{3\sqrt{2}}{\pi} & \frac{4-\sqrt{2}}{\pi}   & \frac{4-2\sqrt{2}}{\pi}
\end{array}
\end{eqnarray}

The corner terms $c_{corn}(\alpha \beta)$ are summarized in the following expression
\begin{equation}
\begin{array}{ccccc}
   & \pm & 0 & a & b \\
\pm  & \frac{3+\sqrt{2}}{\pi}  &  \frac{3\sqrt{2}}{2\pi} &  \frac{1}{\pi}  &   \frac{1-\sqrt{2}}{2\pi}  \\
0  &                      & \frac{3}{\pi}  &  \frac{3\sqrt{2}}{2\pi}  &    \frac{3\sqrt{2}-1}{2\pi}  \\
a  &                      &                        &  \frac{3-\sqrt{2}}{\pi} &    \frac{7-3\sqrt{2}}{2\pi}  \\
b  &                      &                        &                           &   \frac{17-8\sqrt{2}}{4\pi}
\end{array}
\end{equation}
where $\alpha$ is shown in the left column and $\beta$ is shown in the first row. The corner term $c_{corn}(+-)$ is given by
\begin{equation}
c_{corn}(+-)=\frac{3+\sqrt{2}}{\pi}.
\end{equation}
It should be noted that $c_{corn}(+-)=c_{corn}(++)$ as compared with $f_{corn}(+-)\neq f_{corn}(++)$.
The detailed fittings of the numerical data extracting these coefficients are presented in section V.

We also studied the more complicated BCS such as $(+0-0),(+-00),(+a-a)$, etc.
All the results satisfy the expansion form in Eqs. (\ref{eq:free}), (\ref{eq:inter}) and (\ref{eq:specific}). However these results are not presented in this paper.

Let us to summarize the results and give some remarks:

1. The edge and corner logarithmic terms are additive, for example for the surface free energy, $f_c=\sum_{i=1}^{4}f_{corn}(\alpha_i\beta_i) $.

2. There is only one logarithmic correction, the corner term, in the free energy. There are only two logarithmic corrections, the edge and corner terms, in the internal energy. There are only three logarithmic corrections, the leading term, the edge and corner terms in the specific heat. The logarithmic terms are usually related to the singular part of free energy \cite{privman}. Therefore, our results are helpful to determine the singular part of the free energy, which plays a key role in the RG and scaling theory.

3. As we can see, the role of BCs $0,a,b$ is the same in the corner terms for the free energy, but  different in the internal energy and specific heat. In reference \cite{vicari},  using conformal-field theory, the authors classify the possible irrelevant operators for the Ising model with periodic BCs. Now we can ask a similar question: for the BCs $\pm,0,a,b$, what boundary states exist? What irrelevant operators cause the differences in the internal energy and specific heat for these BCs?

\section{The critical free energy density}

\begin{table}
 \caption{ The fitted parameters for the free energy density on with BCs (++++). It has $\delta_{max}<10^{-29}$. }
\begin{tabular}{crr}

\hline
            &  $f_i$~~~~~~~~~~~~~~~~~~~~~~~~~~~~~      &  $\Delta f_i$ ~~~~\\
\hline
$f_0 $  &   $-$0.929695398341610214985384974048D+00&     0.2D$-$27 \\
$f_s $  &   $-$0.802780782154437613604033237415D+00&     0.1D$-$23 \\
$f_c $  &   $-$0.125000000000000000003422789284D+00&     0.1D$-$20 \\
$f_2 $  &    0.659454133542160439126155328533D$-$01&     0.1D$-$19 \\
$f_3 $  &   $-$0.213388347648318443225906415699D+00&     0.1D$-$17 \\
$f_4 $  &    0.168424610272877278595291741255D+00&     0.2D$-$15 \\
$f_5 $  &   $-$0.141070624064625836290050559848D+00&     0.4D$-$13 \\
$f_6 $  &    0.331665779521174569869008623271D$-$01&     0.6D$-$11 \\
$f_7 $  &    0.394787164071544208478676932729D+00&     0.8D$-$09 \\
$f_8 $  &   $-$0.195066002462962494902861949904D+01&     0.7D$-$07 \\
$f_9 $  &    0.746836564272087646811003230908D+01&     0.5D$-$05 \\
$f_{10}$  &   $-$0.271665803915973212020244842054D+02&     0.2D$-$03 \\
$f_{11}$  &    0.995208830272022487362216276397D+02&     0.8D$-$02 \\
$f_{12}$  &   $-$0.375678160708042054486460650382D+03&     0.2D+00 \\
$f_{13}$  &    0.143823385656100850937146960998D+04&     0.4D+01 \\
$f_{14}$  &   $-$0.497443746660492257440708724890D+04&     0.4D+02 \\
$f_{15}$  &    0.109441220818572222121229405839D+05&     0.2D+03 \\

\hline
\end{tabular}
\label{tab:free-++++}
\end{table}
In the numerical calculation for all cases, the number of data points are more than $80$ and the fitting interval of sizes is $30 < N < 1,000$. For the BCs $(aaaa)$, $(+a+a)$, $(a0a0)$, the lattice size is assigned to be $N=2m$, where $m$ is an integer considering the unit length is $2$ for the ``$a$" type boundary. For the BCs $(bbbb)$, $(+b+b)$, $(abab)$ and $(b0b0)$, the lattice size is assigned to be $N=4m$ considering the unit length is $4$ for the ``$b$" type boundary.

To characterize the accuracy of our fittings, we define the maximal deviation
\begin{equation}
\delta_{max}=Max{|y_i-y_i^{fit}|},
\label{eq:error}
\end{equation}
 where $y_i$ is the numerical data and $y_i^{fit}$ is the value given by the fitting formula. We choose the maximum
of the deviations from the data to the fitted ones to represent our fitting quality. We expand the the free energy, internal energy and specific heat to as high order as possible to make the $\delta_{max}$ as small as possible. In every table of the fitting parameters, we give the maximum of deviation. For example, in Tab. \ref{tab:free-++++}, $\delta_{max}=10^{-29}$. We list 30 digits for these results for the convenience of future readers who may want to check other functional forms for the size dependence of the free energy.

The free energy density is fitted with the following formula
\begin{equation}
f=f_0+\frac{f_s}{N} +
\frac{f_c\ln N }{N^2} +\sum_{k=2}^{k_{max}}\frac{f_k}{N^k}
\label{free}
\end{equation}
where $f_s=\sum_{i=1}^{4}f_{\rm surf}(\alpha_i)$ is the sum of the four edges' contribution, and $f_c=\sum_{i=1}^4f_{\rm corn}(\alpha_i\beta_i)$ is the sum of the four corners' contribution. In all fittings of the free energy, we take $k_{max}=15$.

The bulk term is known from the Onsager's exact result \cite{onsager}
\begin{eqnarray}
f_0 & =  & -\ln  \sqrt{2}-\frac{2}{\pi}G \nonumber \\
           & = & -0.9296953983416102149853849736 \cdots
\label{f-bulk}
\end{eqnarray}
where $G=1-\frac{1}{3^2}+\frac{1}{5^2}-\frac{1}{7^2}+\cdots$. This bulk term can be the benchmark for our numerical results.

For the BCs $(++++)$ as shown in Tab. \ref{tab:free-++++}, the $f_0$ agree with the exact result in Eq. (\ref{f-bulk}) in accuracy $10^{-27}$. Assuming the surface term is $f_s=  4 f_{surf}(+)$, we get
\begin{equation}
f_{surf}(+)=-0.200695195538609403401008(3)
\label{eq:surf+}
\end{equation}
Similarly assuming the corner term $f_c=4f_{corn}(++)$ we find that it is satisfied
\begin{equation}
 f_{corn}(++)=-\frac{1}{32}
\end{equation}
in the accuracy of $10^{-20}$. This agrees with the CFT discussion in Eq. (\ref{eq:cft}).

For the BCs $(0000)$, it has been studied in the previous work \cite{wu2,wu3}. The surface term is given  by
\begin{eqnarray}
f_{surf}(0)& = & \frac{1}{2}[\frac{1}{2}\ln(1+\sqrt{2})-D_1] \nonumber \\
           & = &0.09086570849220937845\cdots ,
\label{eq:surf0}
\end{eqnarray}
The corner term $f_{corn}(00)$ has been studied in the previous work \cite{wu2} and it is given by
\begin{equation}
 f_{corn}(00)=-\frac{1}{32}
\end{equation}
which is predicted by the CFT  \cite{cardy}.

\begin{table}
 \caption{ The fitted parameters for the free energy density with BCs ($aaaa$). It has $\delta_{max}<10^{-29}$.}
\begin{tabular}{crr}

\hline
            &  $f_i$~~~~~~~~~~~~~~~~~~~~~~~~~~~~~      &  $\Delta f_i$ ~~~~\\
\hline
$f_0 $  &   $-$0.929695398341610214985384988549D+00&     0.5D$-$26 \\
$f_s $  &    0.233040723946973422962282955610D+00&     0.3D$-$22 \\
$f_c $  &   $-$0.125000000000000000157055887207D+00&     0.4D$-$19 \\
$f_2 $  &   $-$0.254648219025040760355964102628D+00&     0.3D$-$18 \\
$f_3 $  &   $-$0.366116523516816947147615755151D$-$01&     0.3D$-$16 \\
$f_4 $  &    0.990726933828454115319989365025D$-$01&     0.7D$-$14 \\
$f_5 $  &   $-$0.753357380321163693773307018417D$-$01&     0.2D$-$11 \\
$f_6 $  &   $-$0.104607942787190318628716731538D+01&     0.3D$-$09 \\
$f_7 $  &    0.105534696236377112067374836870D+01&     0.4D$-$07 \\
$f_8 $  &   $-$0.597179295600902968567989446877D+01&     0.4D$-$05 \\
$f_9 $  &    0.272745693594755073498173395278D+01&     0.3D$-$03 \\
$f_{10}$  &   $-$0.634658223681635018207456314726D+02&     0.2D$-$01 \\
$f_{11}$  &   $-$0.308992903483955605753873979369D+02&     0.6D+00 \\
$f_{12}$  &   $-$0.108464625658133750579932251299D+04&     0.2D+02 \\
$f_{13}$  &   $-$0.115812044022294120235130937169D+05&     0.4D+03 \\
$f_{14}$  &    0.105846600469528016016531790016D+06&     0.4D+04 \\
$f_{15}$  &   $-$0.143015940838050396032921863125D+07&     0.2D+05 \\

\hline
\end{tabular}
\label{tab:free-aaaa}
\end{table}

\begin{table}
 \caption{ The fitted parameters for the free energy density with BCs ($bbbb$). It has $\delta_{max}<10^{-26}$. }
\begin{tabular}{crr}

\hline
            &  $f_i$~~~~~~~~~~~~~~~~~~~~~~~~~~~~~      &  $\Delta f_i$ ~~~~\\
\hline

$f_0 $  &   $-$0.929695398341610214985391705826D+00&     0.4D$-$23 \\
$f_s $  &    0.143893988489021386166858916572D+00&     0.3D$-$19 \\
$f_c $  &   $-$0.125000000000000072169563957020D+00&     0.4D$-$16 \\
$f_2 $  &   $-$0.773321428035806383051048962333D+00&     0.3D$-$15 \\
$f_3 $  &    0.758252147241525904957559734346D-02&     0.3D$-$13 \\
$f_4 $  &    0.936777139977722272476222906042D-01&     0.7D$-$11 \\
$f_5 $  &   $-$0.686544753242981936307608522488D-01&     0.2D$-$08 \\
$f_6 $  &   $-$0.581572369285089671131602546639D+01&     0.3D$-$06 \\
$f_7 $  &   $-$0.684765080131595012915341607228D+01&     0.4D$-$04 \\
$f_8 $  &   $-$0.264871208008719011917273420988D+02&     0.4D$-$02 \\
$f_9 $  &   $-$0.200027862654843009696442665461D+03&     0.3D+00 \\
$f_{10}$  &   $-$0.704807485097113064528810108323D+03&     0.2D+02 \\
$f_{11}$  &   $-$0.103702182967400052971483563405D+05&     0.6D+03 \\
$f_{12}$  &    0.130782127858347553561739063922D+06&     0.2D+05 \\
$f_{13}$  &   $-$0.427589918936670520764802931284D+07&     0.4D+06 \\
$f_{14}$  &    0.555673697805125025883673389689D+08&     0.4D+07 \\
$f_{15}$  &   $-$0.454298733974663277153070108973D+09&     0.2D+08 \\

\hline
\end{tabular}
\label{tab:free-bbbb}
\end{table}

From Tab. II for the BCs $(aaaa)$, assuming that $f_s=4f_{surf}(a)$, we get
\begin{equation}
f_{surf}(a)=0.0582601809867433557406(6).
\label{eq:surfa}
\end{equation}
In the exact result of infinitely long strip with Brascamp-Kunz BCs, the edge term from the two edges in the expansion of free energy density is  $f_1=\frac{1}{2}\ln (1+\sqrt{2})-\frac{2}{\pi}G$ \cite{izmailian2002b}. The BCs are periodic in the x
direction; in the y direction, the spins are up (+1) along the upper border of the resulting cylinder and have the alternative values along the lower border of the resulting cylinder. In our terminology, Brascamp-Kunz BCs is that the upper border has ``$+$" BCs and the lower border has ``$a$" BCs.  Then it should have $f_1=f_{surf}(+)+f_{surf}(a)$. From the above two equations, we find that it is satisfied indeed
\begin{eqnarray}
&&f_{surf}(+)+f_{surf}(a) \nonumber \\
& = & -0.1424350145518660476604(6) \nonumber \\
&\approx & \frac{1}{2}\ln (1+\sqrt{2})-\frac{2}{\pi}G \nonumber \\
& = & -0.1424350145518660476604642 \cdots
\end{eqnarray}
In reference \cite{izmailian2002b} the two borders are dealt together. From our results, one can see that they can be dealt separately.

For the BCs shown in Tab. II, from the corner term $f_c$, we get
\begin{equation}
f_{corn}(aa)=-\frac{1}{32}
\end{equation}
where we assume $f_c=4f_{corn}(aa)$. The corner term $f_c$ in Tab. II satisfies this equation in the accuracy of $10^{-19}$. This also agrees with the CFT discussion in Eq. (\ref{eq:cft}).

From Tab. III for the BCs $(bbbb)$, assuming that $f_s=4f_{surf}(b)$, we get
\begin{equation}
f_{surf}(b)=0.03597349712225534654(7).
\label{eq:surfb}
\end{equation}
Assuming $f_c=4f_{corn}(bb)$, $f_c$ satisfies that
\begin{equation}
f_{corn}(+-)=\frac{31}{32}
\end{equation}
in the accuracy of $10^{-16}$.  This agrees with the CFT result in Eq. (\ref{eq:cft}).

\begin{table}
 \caption{ The fitted parameters for the free energy density with BCs ($+-+-$). It has $\delta_{max}<10^{-26}$. }
\begin{tabular}{crr}

\hline
            &  $f_i$~~~~~~~~~~~~~~~~~~~~~~~~~~~~~      &  $\Delta f_i$ ~~~~\\
\hline
$f_0 $  &   $-$0.929695398341610214985387693575D+00&     0.2D$-$23 \\
$f_s $  &   $-$0.802780782154437613586470971581D+00&     0.1D$-$19 \\
$f_c $  &    0.387499999999999997581809268118D+01&     0.2D$-$16 \\
$f_2 $  &   $-$0.286882112143865055999777764801D+01&     0.1D$-$15 \\
$f_3 $  &    0.661503877709785276755384453499D+01&     0.1D$-$13 \\
$f_4 $  &   $-$0.879457105742673910757103135424D+01&     0.2D$-$11 \\
$f_5 $  &    0.216271301309375198758440308984D+02&     0.4D$-$09 \\
$f_6 $  &   $-$0.555531271067370240272937773454D+02&     0.7D$-$07 \\
$f_7 $  &    0.161009967789842128529973006611D+03&     0.8D$-$05 \\
$f_8 $  &   $-$0.574780416274328133700304055511D+03&     0.7D$-$03 \\
$f_9 $  &    0.279515822449545159290816563254D+04&     0.5D$-$01 \\
$f_{10}$  &   $-$0.157685780904177657925215704982D+05&     0.2D+01 \\
$f_{11}$  &    0.950949031728661428717508259376D+05&     0.8D+02 \\
$f_{12}$  &   $-$0.586579011630552971109159130875D+06&     0.2D+04 \\
$f_{13}$  &    0.352883992799744590235015340813D+07&     0.4D+05 \\
$f_{14}$  &   $-$0.173348757349422422243427502577D+08&     0.4D+06 \\
$f_{15}$  &    0.490642634772310517762258965649D+08&     0.2D+07 \\

\hline
\end{tabular}
\label{tab:free-+-+-}
\end{table}

In order to investigate the corner terms for two edges under different BCs, we also studied the BCs $(+-+-)$,$(+0+0)$,$(+a+a)$,$(+b+b)$, $(a0a0)$, $(b0b0)$, and $(abab)$. The parameters of fitting of free energy density  for these BCs are shown in Tab. IV-IX respectively.

For the BCs $(+-+-)$ as shown in Tab. IV, the $f_0$ agree with the exact result in Eq. (\ref{f-bulk}) in the accuracy of $10^{-23}$.
For the surface term, due the symmetry, it should have
\begin{equation}
f_{surf}(-)=f_{surf}(+).
\end{equation}
On one hand, the surface term value in Tab. IV satisfies $f_s\approx 4 f_{surf}(+)$ (see Eq. (\ref{eq:surf+})). On the other hand it should have $f_s=2f_{surf}(+)+2f_{surf}(-)$ for the BCs $(+-+-)$. Therefore we verified that $f_{surf}(+)=f_{surf}(-)$ in this case. For the corner term, assuming $f_c=4f_{corn}(+-)$, we get
\begin{equation}
f_{corn}(+-)=\frac{31}{32}
\end{equation}
in the accuracy of $10^{-16}$. This agrees with the CFT result in Eq. (\ref{eq:cft}).

\begin{table}
 \caption{ The fitted parameters for the free energy density with BCs ($+0+0$). It has $\delta_{max}<10^{-29}$. }
\begin{tabular}{crr}

\hline
            &  $f_i$~~~~~~~~~~~~~~~~~~~~~~~~~~~~~      &  $\Delta f_i$ ~~~~\\
\hline
$f_0 $  &   $-$0.929695398341610214985384979151D+00&     0.2D$-$26 \\
$f_s $  &   $-$0.219658974092800053327231355771D+00&     0.1D$-$22 \\
$f_c $  &    0.374999999999999999951194427335D+00&     0.2D$-$19 \\
$f_2 $  &   $-$0.251962670646866032703016870182D+00&     0.1D$-$18 \\
$f_3 $  &    0.679285385210296361122992564020D+00&     0.1D$-$16 \\
$f_4 $  &   $-$0.787770761886194089232866850423D+00&     0.2D$-$14 \\
$f_5 $  &    0.114041298855894871414919327289D+01&     0.4D$-$12 \\
$f_6 $  &   $-$0.195756094548926620389166895374D+01&     0.7D$-$10 \\
$f_7 $  &    0.474170712806294843708705192306D+01&     0.8D$-$08 \\
$f_8 $  &   $-$0.135615816375142369877704378052D+02&     0.7D$-$06 \\
$f_9 $  &    0.415669990322889390551606164473D+02&     0.5D$-$04 \\
$f_{10}$  &   $-$0.138063044779598383122579968043D+03&     0.2D$-$02 \\
$f_{11}$  &    0.539877721617188775593669604842D+03&     0.8D$-$01 \\
$f_{12}$  &   $-$0.244256523028559605382647808389D+04&     0.2D+01 \\
$f_{13}$  &    0.116933684783462486159320612910D+05&     0.4D+02 \\
$f_{14}$  &   $-$0.486720865872993175064582325108D+05&     0.4D+03 \\
$f_{15}$  &    0.123752910451271786291771980697D+06&     0.2D+04 \\

\hline
\end{tabular}
\label{tab:free-+0+0}
\end{table}

For the BCs $(+0+0)$ as shown in Tab. \ref{tab:free-+0+0}, the $f_0$ agree with the exact result in Eq. (\ref{f-bulk}) in accuracy $10^{-26}$. The surface term satisfies $f_s\approx  2 (f_{surf}(+)+f_{surf}(0))$ (see Eq. (\ref{eq:surf+}) and (\ref{eq:surf0})).
Assuming $f_c=4f_{corn}(+0)$, $f_c$ satisfies that
\begin{equation}
f_{corn}(+0)=\frac{3}{32}
\end{equation}
in the accuracy of $10^{-19}$. This agrees with the CFT discussion in Eq. (\ref{eq:cft}).

\begin{table}
 \caption{ The fitted parameters for the free energy density with BCs ($+a+a$). It has $\delta_{max}<10^{-29}$. }
\begin{tabular}{crr}

\hline
            &  $f_i$~~~~~~~~~~~~~~~~~~~~~~~~~~~~~      &  $\Delta f_i$ ~~~~\\
\hline
$f_0 $  &   $-$0.929695398341610214985384978573D+00&     0.5D$-$26 \\
$f_s $  &   $-$0.284870029103732095320894042639D+00&     0.3D$-$22 \\
$f_c $  &    0.374999999999999999947963881455D+00&     0.4D$-$19 \\
$f_2 $  &   $-$0.230528069852075681873945801591D+00&     0.3D$-$18 \\
$f_3 $  &    0.695489561541278037204978151481D+00&     0.3D$-$16 \\
$f_4 $  &   $-$0.103141858806547819142244836497D+01&     0.7D$-$14 \\
$f_5 $  &    0.166549757640691389363999116709D+01&     0.2D$-$11 \\
$f_6 $  &   $-$0.315666329489744134367710436920D+01&     0.3D$-$09 \\
$f_7 $  &    0.751364556071420027856983095629D+01&     0.4D$-$07 \\
$f_8 $  &   $-$0.211724454317169164364511684613D+02&     0.4D$-$05 \\
$f_9 $  &    0.569285933614902400574612055271D+02&     0.3D$-$03 \\
$f_{10}$  &   $-$0.200894879706556167596666677748D+03&     0.2D$-$01 \\
$f_{11}$  &    0.738465234850498407479359039595D+03&     0.6D+00 \\
$f_{12}$  &   $-$0.370625233162063853539474507949D+04&     0.2D+02 \\
$f_{13}$  &    0.139026951179773562321266688861D+05&     0.4D+03 \\
$f_{14}$  &   $-$0.472647416399960719811956345120D+05&     0.4D+04 \\
$f_{15}$  &   $-$0.148448796252492093078011329716D+06&     0.2D+05 \\

\hline
\end{tabular}
\label{tab:free-+a+a}
\end{table}

For the BCs $(+a+a)$ as shown in Tab. \ref{tab:free-+a+a}, the $f_0$ agree with the exact result in Eq. (\ref{f-bulk}) in accuracy $10^{-26}$. The surface term satisfies $f_s\approx  2 (f_{surf}(+)+f_{surf}(a))$ (see Eq. (\ref{eq:surf+}) and (\ref{eq:surfa})).
Assuming $f_c=4f_{corn}(+a)$, $f_c$ satisfies that
\begin{equation}
f_{corn}(+a)=\frac{3}{32}
\end{equation}
in the accuracy of $10^{-19}$. This agrees with the CFT result in Eq. (\ref{eq:cft}).

\begin{table}
 \caption{ The fitted parameters for the free energy density with BCs ($+b+b$). It has $\delta_{max}<10^{-28}$. }
\begin{tabular}{crr}

\hline
            &  $f_i$~~~~~~~~~~~~~~~~~~~~~~~~~~~~~      &  $\Delta f_i$ ~~~~\\
\hline
$f_0 $  &   $-$0.929695398341610214985385813730D+00&     0.4D$-$24 \\
$f_s $  &   $-$0.329443396832708113736448563927D+00&     0.3D$-$20 \\
$f_c $  &    0.374999999999999991011086593966D+00&     0.4D$-$17 \\
$f_2 $  &   $-$0.230019097408478648756634601006D+00&     0.4D$-$16 \\
$f_3 $  &    0.709320691190351009306010364727D+00&     0.3D$-$14 \\
$f_4 $  &   $-$0.120691073041835944419415296734D+01&     0.7D$-$12 \\
$f_5 $  &    0.175074679956124288800019955056D+01&     0.2D$-$09 \\
$f_6 $  &   $-$0.369406345920046582055132086037D+01&     0.3D$-$07 \\
$f_7 $  &    0.735492034159148073767794048547D+01&     0.4D$-$05 \\
$f_8 $  &   $-$0.220405139127909583771177194810D+02&     0.4D$-$03 \\
$f_9 $  &    0.480938801075341021638811454946D+02&     0.3D$-$01 \\
$f_{10}$  &   $-$0.303077347782971225207247495923D+03&     0.2D+01 \\
$f_{11}$  &   $-$0.359847885906329599565504973931D+03&     0.6D+02 \\
$f_{12}$  &    0.128504892136533622130900355317D+05&     0.2D+04 \\
$f_{13}$  &   $-$0.477108955284822145961201822179D+06&     0.4D+05 \\
$f_{14}$  &    0.649445722586115617389099847285D+07&     0.4D+06 \\
$f_{15}$  &   $-$0.521551126077579668375565890603D+08&     0.2D+07 \\

\hline
\end{tabular}
\label{tab:free-+b+b}
\end{table}
For the BCs $(+a+a)$ as shown in Tab. \ref{tab:free-+b+b}, the $f_0$ agree with the exact result in Eq. (\ref{f-bulk}) in accuracy $10^{-24}$. The surface term satisfies $f_s\approx  2 (f_{surf}(+)+f_{surf}(b))$ (see Eq. (\ref{eq:surf+}) and (\ref{eq:surfb})). From the corner term $f_c$ we get
\begin{equation}
f_{corn}(+b)=\frac{3}{32}
\end{equation}
assuming  $f_c=4f_{corn}(+b)$. This agrees with the CFT discussion in Eq. (\ref{eq:cft}). The $f_c$ in Tab. IV coincides this value in the accuracy of $10^{-17}$.

\begin{table}
 \caption{ The fitted parameters for the free energy density with BCs ($a0a0$). It has $\delta_{max}<10^{-26}$. }
\begin{tabular}{crr}

\hline
            &  $f_i$~~~~~~~~~~~~~~~~~~~~~~~~~~~~~      &  $\Delta f_i$ ~~~~\\
\hline
$f_0 $  &   $-$0.929695398341610214985384981291D+00&     0.5D$-$26 \\
$f_s $  &    0.298251778957905464955893931191D+00&     0.3D$-$22 \\
$f_c $  &   $-$0.125000000000000000080779862576D+00&     0.4D$-$19 \\
$f_2 $  &   $-$0.162549953486831879945908727115D+00&     0.3D$-$18 \\
$f_3 $  &   $-$0.625000000000000696722514954026D-01&     0.3D$-$16 \\
$f_4 $  &    0.105056309403774064305710060928D+00&     0.7D$-$14 \\
$f_5 $  &   $-$0.232353723262690069263633915125D+00&     0.2D$-$11 \\
$f_6 $  &   $-$0.374902765755521977036263762574D+00&     0.3D$-$09 \\
$f_7 $  &    0.154458932635911927966650263926D+00&     0.4D$-$07 \\
$f_8 $  &   $-$0.219933075683671700394841947896D+01&     0.4D$-$05 \\
$f_9 $  &   $-$0.179803696486812551637886519807D+01&     0.3D$-$03 \\
$f_{10}$  &   $-$0.251509403561259400399146373084D+02&     0.2D$-$01 \\
$f_{11}$  &   $-$0.582463937453080337439664406645D+02&     0.6D+00 \\
$f_{12}$  &   $-$0.390174798900757832297239523430D+03&     0.2D+02 \\
$f_{13}$  &   $-$0.708524816533114951767771645322D+04&     0.4D+03 \\
$f_{14}$  &    0.592931942938821400564144490776D+05&     0.4D+04 \\
$f_{15}$  &   $-$0.762189243915292907930560466089D+06&     0.2D+05 \\
\hline
\end{tabular}
\label{tab:free-a0a0}
\end{table}

For the BCs $(a0a0)$ as shown in Tab. \ref{tab:free-a0a0}, the $f_0$ agree with the exact result in Eq. (\ref{f-bulk}) in accuracy $10^{-26}$. The surface term satisfies $f_s\approx  2 (f_{surf}(a)+f_{surf}(0))$ (see Eq. (\ref{eq:surfa}) and (\ref{eq:surf0})).
Assuming $f_c=4f_{corn}(a0)$, $f_c$ satisfies that
\begin{equation}
f_{corn}(a0)=-\frac{1}{32}
\end{equation}
in the accuracy of $10^{-16}$. This agrees with the CFT discussion in Eq. (\ref{eq:cft}).

\begin{table}
 \caption{ The fitted parameters for the free energy density with BCs ($b0b0$). It has $\delta_{max}<10^{-26}$.}
\begin{tabular}{crr}

\hline
            &  $f_i$~~~~~~~~~~~~~~~~~~~~~~~~~~~~~      &  $\Delta f_i$ ~~~~\\
\hline
$f_0 $  &   $-$0.929695398341610214985392172843D+00&     0.4D$-$23 \\
$f_s $  &    0.253678411228929446585274984858D+00&     0.3D$-$19 \\
$f_c $  &   $-$0.125000000000000077176293005476D+00&     0.4D$-$16 \\
$f_2 $  &   $-$0.335634413581536397718566158572D+00&     0.3D$-$15 \\
$f_3 $  &   $-$0.404029130879871203000598790501D-01&     0.3D$-$13 \\
$f_4 $  &    0.359798387750386260241926679420D+00&     0.7D$-$11 \\
$f_5 $  &   $-$0.116703428358044570379267432362D+01&     0.2D$-$08 \\
$f_6 $  &   $-$0.332883620213668316652053277123D+01&     0.3D$-$06 \\
$f_7 $  &   $-$0.422759697240003310703707474918D+01&     0.4D$-$04 \\
$f_8 $  &   $-$0.264827721482098275372874648284D+02&     0.4D$-$02 \\
$f_9 $  &   $-$0.150582257563333053261020216199D+03&     0.3D+00 \\
$f_{10}$  &   $-$0.733611959412881429322346117682D+03&     0.2D+02 \\
$f_{11}$  &   $-$0.915492982933561739818861525106D+04&     0.6D+03 \\
$f_{12}$  &    0.137615187296902941392679011802D+06&     0.2D+05 \\
$f_{13}$  &   $-$0.453409486598948067430631596683D+07&     0.4D+06 \\
$f_{14}$  &    0.594917177554898718315162351265D+08&     0.4D+07 \\
$f_{15}$  &   $-$0.481662437905079056507669248723D+09&     0.2D+08 \\

\hline
\end{tabular}
\label{tab:free-b0b0}
\end{table}

For the BCs $(a0a0)$ as shown in Tab. \ref{tab:free-b0b0}, the $f_0$ agree with the exact result in Eq. (\ref{f-bulk}) in accuracy $10^{-23}$. The surface term satisfies $f_s\approx  2 (f_{surf}(b)+f_{surf}(0))$ (see Eq. (\ref{eq:surfb}) and (\ref{eq:surf0})).
Assuming $f_c=4f_{corn}(b0)$, $f_c$ satisfies that
\begin{equation}
f_{corn}(b0)=-\frac{1}{32}
\end{equation}
in the accuracy of $10^{-16}$. This also agrees with the CFT result in Eq. (\ref{eq:cft}).

\begin{table}
 \caption{ The fitted parameters for the free energy density with BCs ($abab$). It has $\delta_{max}<10^{-26}$.}
\begin{tabular}{crr}

\hline
            &  $f_i$~~~~~~~~~~~~~~~~~~~~~~~~~~~~~      &  $\Delta f_i$ ~~~~\\
\hline
$f_0 $  &   $-$0.929695398341610214985391063571D+00&     0.4D$-$23 \\
$f_s $  &    0.188467356217997404583768325312D+00&     0.3D$-$19 \\
$f_c $  &   $-$0.125000000000000065271195810734D+00&     0.4D$-$16 \\
$f_2 $  &   $-$0.402044240185024616285674357904D+00&     0.3D$-$15 \\
$f_3 $  &   $-$0.145145654396583782915730330964D-01&     0.3D$-$13 \\
$f_4 $  &    0.234254986629528683380199225763D+00&     0.7D$-$11 \\
$f_5 $  &   $-$0.321686999853399366938884218978D+00&     0.2D$-$08 \\
$f_6 $  &   $-$0.372433958605540485789211548474D+01&     0.3D$-$06 \\
$f_7 $  &   $-$0.165719338242722840573697946386D+01&     0.4D$-$04 \\
$f_8 $  &   $-$0.248175457151780509309853663343D+02&     0.4D$-$02 \\
$f_9 $  &   $-$0.108550900091291780039578332436D+03&     0.3D+00 \\
$f_{10}$  &   $-$0.679716727388865517955336704321D+03&     0.2D+02 \\
$f_{11}$  &   $-$0.753209986254719150087212966579D+04&     0.6D+03 \\
$f_{12}$  &    0.117378369628757425304490851691D+06&     0.2D+05 \\
$f_{13}$  &   $-$0.379862636122939743872519633388D+07&     0.4D+06 \\
$f_{14}$  &    0.499820018018333394100724421053D+08&     0.4D+07 \\
$f_{15}$  &   $-$0.403602527784390307629146911536D+09&     0.2D+08 \\

\hline
\end{tabular}
\label{tab:free-abab}
\end{table}

For the BCs $(abab)$ as shown in Tab. \ref{tab:free-abab}, the $f_0$ agree with the exact result in Eq. (\ref{f-bulk}) in accuracy $10^{-23}$. The surface term satisfies $f_s\approx  2 (f_{surf}(a)+f_{surf}(b))$ (see Eq. (\ref{eq:surfa}) and (\ref{eq:surfb})).
Assuming $f_c=4f_{corn}(ab)$, $f_c$ satisfies that
\begin{equation}
f_{corn}(ab)=-\frac{1}{32}
\end{equation}
in the accuracy of $10^{-16}$. This agrees with the CFT result in Eq. (\ref{eq:cft}).

\section{The critical internal energy density}

We find that the critical internal energy  can be expanded into
\begin{equation}
u=u_0+u_{\rm surf}\frac{\ln N}{N} +u_{\rm corn}
\frac{\ln N }{N^2} +\sum_{k=1}^{\infty}\frac{B_k}{N^k}.
\label{squ-inter1}
\end{equation}
where the surface term $u_s=\sum_{i=1}^{4}u_{\rm surf}(\alpha_i)$ is the sum of the four edges' contribution and the corner term $u_c=\sum_{i=1}^4u_{\rm corn}(\alpha_i\beta_i)$ is the sum of the four corners' contribution.

For all BCs, there should have the same bulk internal energy $u_0$, which is obtained in the Onsager's exact solution \cite{onsager}
\begin{equation}
u_0= -\sqrt{2}=-1.41421356237309504880168872\cdots
\end{equation}
This is the benchmark for our fittings.

For the free BCs $(0000)$, the edge term $u_{surf}(0)$ and $U_{corn}(0)$ have been obtained in the previous work \cite{wu}. According to the definition of this paper, they are given by
\begin{equation}
u_{surf}(0)=\frac{1}{\pi}
\label{eq:usurf-0}
\end{equation}
and
\begin{equation}
u_{corn}(00)=\frac{\sqrt{2}}{2\pi}.
\end{equation}

\begin{table}
 \caption{ The fitted parameters for the internal energy density with BCs (++++). It has $\delta_{max}<10^{-29}$.}
\begin{tabular}{crr}

\hline
            &  $u_i$~~~~~~~~~~~~~~~~~~~~~~~~~~~~~      &  $\Delta u_i$ ~~~~\\
\hline
$u_0 $  &   $-$0.141421356237309504880168869009D+01&     0.4D$-$25 \\
$u_s $  &   $-$0.127323954473516268615128913161D+01&     0.2D$-$21 \\
$u_1 $  &   $-$0.121362669205892989034286995905D+00&     0.2D$-$20 \\
$u_c $  &   $-$0.217355586089226875600888009916D+01&     0.3D$-$18 \\
$u_2 $  &   $-$0.966521334103453417350504802697D+00&     0.2D$-$17 \\
$u_3 $  &   $-$0.183246526200041716524194321901D+01&     0.9D$-$16 \\
$u_4 $  &    0.896206745066539545305761229339D+00&     0.1D$-$13 \\
$u_5 $  &   $-$0.370138800180383773990721272130D+00&     0.2D$-$11 \\
$u_6 $  &   $-$0.759739103052453377282182330013D+00&     0.3D$-$09 \\
$u_7 $  &    0.433694612591844987383362568350D+01&     0.3D$-$07 \\
$u_8 $  &   $-$0.163237649671144636075513812223D+02&     0.2D$-$05 \\
$u_9 $  &    0.573226743976490532699177503300D+02&     0.1D$-$03 \\
$u_{10}$  &   $-$0.202923678621435723037876791004D+03&     0.6D$-$02 \\
$u_{11}$  &    0.749578627676759522362969270939D+03&     0.2D+00 \\
$u_{12}$  &   $-$0.293163088715786209592251230669D+04&     0.5D+01 \\
$u_{13}$  &    0.117770961965249336674404764102D+05&     0.8D+02 \\
$u_{14}$  &   $-$0.424904751403293739212434528571D+05&     0.8D+03 \\
$u_{15}$  &    0.956249625516078094156574902045D+05&     0.4D+04 \\
\hline
\end{tabular}
\label{Tab:energy-++++}
\end{table}

For the BCs $(++++)$, the fitting parameters are shown in Tab. \ref{Tab:energy-++++}. We confirm the exact result for the bulk internal energy density in accuracy of $10^{-25}$. Assuming $u_s=4u_{surf}(+)$, we conjecture
\begin{equation}
u_{surf}(+)=-\frac{1}{\pi}
\label{eq:usurf-+}
\end{equation}
The corner term $u_s$ in Tab. \ref{Tab:energy-++++} agrees with this conjecture in the accuracy of $10^{-21}$.
For the corner term, assuming $u_c=4u_{corn}(++)$, we conjecture that
\begin{equation}
u_{corn}(++)=-\frac{2+\sqrt{2}}{2\pi}
\end{equation}
This is valid in the accuracy of $10^{-18}$.

\begin{table}
 \caption{ The fitted parameters for the internal energy density with BCs ($aaaa$). It has $\delta_{max}<10^{-29}$. }
\begin{tabular}{crr}

\hline
            &  $u_i$~~~~~~~~~~~~~~~~~~~~~~~~~~~~~      &  $\Delta u_i$ ~~~~\\
\hline
$u_0 $  &   $-$0.141421356237309504880168832528D+01&     0.8D$-$25 \\
$u_s $  &    0.127323954473516268614827101431D+01&     0.5D$-$21 \\
$u_1 $  &    0.121362669205892989060432127750D+00&     0.5D$-$20 \\
$u_c $  &    0.372923228578056612338835402163D+00&     0.7D$-$18 \\
$u_2 $  &    0.165828844286277761671386974606D+00&     0.4D$-$17 \\
$u_3 $  &    0.436315136854690268228847146263D+00&     0.3D$-$15 \\
$u_4 $  &   $-$0.237734966941290106536283662171D+00&     0.5D$-$13 \\
$u_5 $  &   $-$0.144345615308223445290429547041D+01&     0.8D$-$11 \\
$u_6 $  &    0.679132792426142902337123922004D+00&     0.1D$-$08 \\
$u_7 $  &   $-$0.718330820940823459196862516371D+01&     0.1D$-$06 \\
$u_8 $  &    0.361448823835112932181123936506D+01&     0.1D$-$04 \\
$u_9 $  &   $-$0.488626495203784700233871256267D+02&     0.8D$-$03 \\
$u_{10}$  &   $-$0.980553971839542955609196693808D+01&     0.4D$-$01 \\
$u_{11}$  &   $-$0.897861066154329812750995018843D+03&     0.2D+01 \\
$u_{12}$  &    0.426690691676072366467068550187D+03&     0.5D+02 \\
$u_{13}$  &   $-$0.559614222557187177133582038801D+05&     0.8D+03 \\
$u_{14}$  &    0.432377686667277475645947156909D+06&     0.9D+04 \\
$u_{15}$  &   $-$0.471174327437116160296769294951D+07&     0.5D+05 \\

\hline
\end{tabular}
\label{Tab:energy-aaaa}
\end{table}

Tab. \ref{Tab:energy-aaaa} is for the BCs $(aaaa)$. The fitted bulk term $u_0$  confirm the exact result in accuracy of $10^{-25}$. Assuming $u_s=4u_{surf}(a)$, we conjecture
\begin{equation}
u_{surf}(a)=\frac{1}{\pi}
\label{eq:usurf-a}
\end{equation}
The corner term $u_s$ in Tab. \ref{Tab:energy-aaaa} agrees with this conjecture in the accuracy of $10^{-21}$.
For the corner term, assuming $u_c=4u_{corn}(aa)$, we conjecture that
\begin{equation}
u_{corn}(aa)=\frac{2-\sqrt{2}}{2\pi}
\end{equation}
This is valid in the accuracy of $10^{-18}$ (see Tab.\ref{Tab:energy-aaaa}).

\begin{table}
 \caption{ The fitted parameters for the internal energy density with BCs ($bbbb$). It has $\delta_{max}<10^{-26}$.}
\begin{tabular}{crr}

\hline
            &  $u_i$~~~~~~~~~~~~~~~~~~~~~~~~~~~~~      &  $\Delta u_i$ ~~~~\\
\hline
$u_0 $  &   $-$0.141421356237309504880135950588D+01&     0.8D$-$22 \\
$u_s $  &    0.127323954473516268382279793781D+01&     0.5D$-$18 \\
$u_1 $  &    0.509820355726882837098240237571D-01&     0.4D$-$17 \\
$u_c $  &   $-$0.772349295004999797295679402263D-01&     0.7D$-$15 \\
$u_2 $  &   $-$0.280722865969988752004150305887D+01&     0.4D$-$14 \\
$u_3 $  &    0.645371814938644881713223447427D+00&     0.3D$-$12 \\
$u_4 $  &    0.118989316191903623819336234630D+00&     0.5D$-$10 \\
$u_5 $  &   $-$0.900771716580585293775278759690D+01&     0.8D$-$08 \\
$u_6 $  &   $-$0.127866839483116151516218830460D+02&     0.1D$-$05 \\
$u_7 $  &   $-$0.446239252197777558682759751167D+02&     0.1D$-$03 \\
$u_8 $  &   $-$0.208032124249502693647501621960D+03&     0.1D$-$01 \\
$u_9 $  &   $-$0.719157476367560572470304735699D+03&     0.8D+00 \\
$u_{10}$  &   $-$0.426141888628467713474610491665D+04&     0.4D+02 \\
$u_{11}$  &   $-$0.525605460374128169653367198990D+05&     0.2D+04 \\
$u_{12}$  &    0.814226240024869799104869638369D+06&     0.5D+05 \\
$u_{13}$  &   $-$0.253642335953491210461671657992D+08&     0.8D+06 \\
$u_{14}$  &    0.324881319142663462520385239658D+09&     0.9D+07 \\
$u_{15}$  &   $-$0.253360729395602388584637091373D+10&     0.5D+08 \\

\hline
\end{tabular}
\label{Tab:energy-bbbb}
\end{table}

Tab. \ref{Tab:energy-bbbb} is for the BCs $(bbbb)$. The fitted bulk term $u_0$  confirm the exact result for the bulk internal energy density in accuracy of $10^{-22}$. Assuming $u_s=4u_{surf}(b)$, we conjecture
\begin{equation}
u_{surf}(b)=\frac{1}{\pi}
\label{eq:usurf-b}
\end{equation}
The corner term $u_s$ in Tab. \ref{Tab:energy-bbbb} agrees with this conjecture in the accuracy of $10^{-18}$.
For the corner term, assuming $u_c=4u_{corn}(bb)$, we conjecture that
\begin{equation}
u_{corn}(bb)=\frac{4-3\sqrt{2}}{4\pi}
\end{equation}
This is valid in the accuracy of $10^{-15}$ (see Tab.\ref{Tab:energy-bbbb})

\begin{table}
 \caption{ The fitted parameters for the internal energy density with BCs ($+-+-$). It has $\delta_{max}<10^{-25}$. }
\begin{tabular}{crr}

\hline
            &  $u_i$~~~~~~~~~~~~~~~~~~~~~~~~~~~~~      &  $\Delta u_i$ ~~~~\\
\hline
$u_0 $  &   $-$0.141421356237309504880209105160D+01&     0.4D$-$21 \\
$u_s $  &   $-$0.127323954473516268355937634327D+01&     0.2D$-$17 \\
$u_1 $  &    0.592933381617406067698112929821D+01&     0.2D$-$16 \\
$u_c $  &   $-$0.217355586089226516559356301131D+01&     0.3D$-$14 \\
$u_2 $  &   $-$0.429419058250212614263404544988D+01&     0.2D$-$13 \\
$u_3 $  &   $-$0.531102203261029549523581183270D+01&     0.9D$-$12 \\
$u_4 $  &    0.145823238839076108926623938355D+02&     0.1D$-$09 \\
$u_5 $  &   $-$0.283124145434214014731214582951D+02&     0.2D$-$07 \\
$u_6 $  &    0.958963815165193163683300790488D+02&     0.3D$-$05 \\
$u_7 $  &   $-$0.497036792141350475751154076725D+03&     0.3D$-$03 \\
$u_8 $  &    0.348367843810626358923766114664D+04&     0.2D$-$01 \\
$u_9 $  &   $-$0.225412512712345257960614002739D+05&     0.1D+01 \\
$u_{10}$  &    0.145416996000500434212352640009D+06&     0.6D+02 \\
$u_{11}$  &   $-$0.960622291662460239451060933181D+06&     0.2D+04 \\
$u_{12}$  &    0.665451316615658302773072238739D+07&     0.5D+05 \\
$u_{13}$  &   $-$0.430512534233410462587392770433D+08&     0.8D+06 \\
$u_{14}$  &    0.218919044042652809955014163869D+09&     0.8D+07 \\
$u_{15}$  &   $-$0.590365649155689022900464405037D+09&     0.4D+08 \\

\hline
\end{tabular}
\label{Tab:energy-+-+-}
\end{table}
As shown in Tab. \ref{Tab:energy-+-+-} for BCs $(+-+-)$, we confirm the exact result for the bulk internal energy density in accuracy of $10^{-21}$. Assuming $u_s=2u_{surf}(+)+2u_{surf}(-)$ and using Eq. (\ref{eq:usurf-+}) , we get
\begin{equation}
u_{surf}(-)=-\frac{1}{\pi}
\label{eq:usurf--}
\end{equation}
in an accuracy $10^{-17}$. The symmetry requires that $u_{surf}(-)=u_{surf}(+)$. The edge term $u_s$ in Tab. \ref{Tab:energy-+-+-} agrees with this conjecture in the accuracy of $10^{-17}$.

For the corner term, assuming $u_c=4u_{corn}(+-)$, we conjecture that
\begin{equation}
u_{corn}(+-)=-\frac{2+\sqrt{2}}{2\pi}=u_{corn}(++)
\end{equation}
This is valid in the accuracy of $10^{-14}$.

\begin{table}
 \caption{ The fitted parameters for the internal energy density with BCs ($+0+0$). It has $\delta_{max}<10^{-29}$. }
\begin{tabular}{crr}

\hline
            &  $u_i$~~~~~~~~~~~~~~~~~~~~~~~~~~~~~      &  $\Delta u_i$ ~~~~\\
\hline
$u_0 $  &   $-$0.141421356237309504880168854931D+01&     0.4D$-$25 \\
$u_s $  &   $-$0.112593854595141369838108423955D-20&     0.2D$-$21 \\
$u_1 $  &   $-$0.236420159158849830728148914245D+00&     0.2D$-$20 \\
$u_c $  &    0.636619772367581341517422861073D+00&     0.3D$-$18 \\
$u_2 $  &   $-$0.403980179740954679523416249621D+00&     0.2D$-$17 \\
$u_3 $  &    0.153039004599883637932465360730D+01&     0.9D$-$16 \\
$u_4 $  &   $-$0.224546982818408767553909046449D+01&     0.1D$-$13 \\
$u_5 $  &    0.312569900114974929426200166902D+01&     0.2D$-$11 \\
$u_6 $  &   $-$0.575105607247354754587930290964D+01&     0.3D$-$09 \\
$u_7 $  &    0.167891484006176393416531025223D+02&     0.3D$-$07 \\
$u_8 $  &   $-$0.518607584028372718460773951224D+02&     0.2D$-$05 \\
$u_9 $  &    0.148353618151306805204141990816D+03&     0.1D$-$03 \\
$u_{10}$  &   $-$0.452385710143795259985560797555D+03&     0.6D$-$02 \\
$u_{11}$  &    0.172748958195878547531727368943D+04&     0.2D+00 \\
$u_{12}$  &   $-$0.801383326507455157968258145447D+04&     0.50D+01 \\
$u_{13}$  &    0.375697090623553345562944854271D+05&     0.8D+02 \\
$u_{14}$  &   $-$0.150209491850475103454220970397D+06&     0.8D+03 \\
$u_{15}$  &    0.344351474160398620172396213247D+06&     0.4D+04 \\

\hline
\end{tabular}
\label{Tab:energy-+0+0}
\end{table}

Tab. \ref{Tab:energy-+0+0} is for the BCs $(+0+0)$. The fitted bulk term $u_0$  agrees with the exact result  in accuracy of $10^{-25}$. Assuming $u_s=2u_{surf}(+)+2u_{surf}(0)$, we should have $u_s=0$ following Eq. (\ref{eq:usurf-+}) and (\ref{eq:usurf-0}). The surface term in Tab. \ref{Tab:energy-+0+0} $|u_s| < 10^{-20}$ is nearly zero. For the corner term, assuming $u_c=4u_{corn}(+0)$, we conjecture that
\begin{equation}
u_{corn}(+0)=\frac{1}{2\pi}
\end{equation}
This is valid in the accuracy of $10^{-18}$ (see Tab.\ref{Tab:energy-+0+0})

\begin{table}
 \caption{ The fitted parameters for the internal energy density with BCs ($+a+a$). It has $\delta_{max}<10^{-29}$. }
\begin{tabular}{crr}

\hline
            &  $u_i$~~~~~~~~~~~~~~~~~~~~~~~~~~~~~      &  $\Delta u_i$ ~~~~\\
\hline
$u_0 $  &   $-$0.141421356237309504880168906269D+01&     0.8D$-$25 \\
$u_s $  &    0.237281950923559599515533308346D-20&     0.5D$-$21 \\
$u_1 $  &   $-$0.236420159158849830758376907561D+00&     0.5D$-$20 \\
$u_c $  &    0.900316316157106073159541773205D+00&     0.7D$-$18 \\
$u_2 $  &   $-$0.818353513742248276170560645526D+00&     0.4D$-$17 \\
$u_3 $  &    0.247965110422734714389898659784D+01&     0.3D$-$15 \\
$u_4 $  &   $-$0.368355719504005167143843674343D+01&     0.5D$-$13 \\
$u_5 $  &    0.423781573788511073622571525775D+01&     0.8D$-$11 \\
$u_6 $  &   $-$0.774487104335495173537469831101D+01&     0.1D$-$08 \\
$u_7 $  &    0.261717126843229249248555559278D+02&     0.1D$-$06 \\
$u_8 $  &   $-$0.730212428625000966547392691302D+02&     0.1D$-$04 \\
$u_9 $  &    0.190936147297823282092652988992D+03&     0.8D$-$03 \\
$u_{10}$  &   $-$0.541080946525900119542013065860D+03&     0.4D$-$01 \\
$u_{11}$  &    0.259773209516561793637165819874D+04&     0.2D+01 \\
$u_{12}$  &   $-$0.113206201546654466041419051873D+05&     0.5D+02 \\
$u_{13}$  &    0.910179790390602075667908122548D+05&     0.8D+03 \\
$u_{14}$  &   $-$0.611557790179637302459722595349D+06&     0.9D+04 \\
$u_{15}$  &    0.447394715623069310269646669090D+07&     0.5D+05 \\

\hline
\end{tabular}
\label{Tab:energy-+a+a}
\end{table}

Tab. \ref{Tab:energy-+a+a} is for the BCs $(+a+a)$. The fitted bulk term $u_0$  agrees with the exact result $u_0=\sqrt{2}$ in the accuracy of $10^{-25}$. Assuming $u_s=2u_{surf}(+)+2u_{surf}(a)$, we should have $u_s=0$ following Eq. (\ref{eq:usurf-+}) and (\ref{eq:usurf-b}). The surface term in Tab. \ref{Tab:energy-+a+a} $|u_s| < 10^{-20}$ is nearly zero indeed. For the corner term, assuming $u_c=4u_{corn}(+a)$, we conjecture that
\begin{equation}
u_{corn}(+a)=\frac{\sqrt{2}}{2\pi}
\end{equation}
This is valid in the accuracy of $10^{-15}$ (see Tab.\ref{Tab:energy-+a+a})

\begin{table}
 \caption{ The fitted parameters for the internal energy density with BCs ($+b+b$). It has $\delta_{max}<10^{-26}$.}
\begin{tabular}{crr}

\hline
            &  $u_i$~~~~~~~~~~~~~~~~~~~~~~~~~~~~~      &  $\Delta u_i$ ~~~~\\
\hline
$u_0 $  &   $-$0.141421356237309504880189175368D+01&     0.8D$-$22 \\
$u_s $  &    0.143287214304804212911925468018D-17&     0.5D$-$18 \\
$u_1 $  &   $-$0.271610475975452205930154024922D+00&     0.4D$-$17 \\
$u_c $  &    0.112539539519638477334064523305D+01&     0.7D$-$15 \\
$u_2 $  &   $-$0.110255538722893398035041997049D+01&     0.4D$-$14 \\
$u_3 $  &    0.311422905225047272143244029409D+01&     0.3D$-$12 \\
$u_4 $  &   $-$0.443239467881127088850654545434D+01&     0.5D$-$10 \\
$u_5 $  &    0.203930544807966386515444358254D+01&     0.8D$-$08 \\
$u_6 $  &   $-$0.810394378580850081275885908197D+01&     0.1D$-$05 \\
$u_7 $  &    0.305088519713148254145725680487D+02&     0.1D$-$03 \\
$u_8 $  &   $-$0.574901549849639364714542984031D+02&     0.1D$-$01 \\
$u_9 $  &    0.240732545248500143568194492225D+03&     0.8D+00 \\
$u_{10}$  &   $-$0.208587999021048046958271025162D+03&     0.4D+02 \\
$u_{11}$  &    0.246346842936457817928191776432D+05&     0.2D+04 \\
$u_{12}$  &   $-$0.557338643025096581803585979886D+06&     0.5D+05 \\
$u_{13}$  &    0.135456580934878293298693779823D+08&     0.8D+06 \\
$u_{14}$  &   $-$0.178326622663623585761256595815D+09&     0.9D+07 \\
$u_{15}$  &    0.125499772276245750941179598653D+10&     0.5D+08 \\

\hline
\end{tabular}
\label{Tab:energy-+b+b}
\end{table}

Tab. \ref{Tab:energy-+b+b} is for the BCs $(+b+b)$. The fitted bulk term $u_0$  agrees with the exact result $u_0=\sqrt{2}$ in the accuracy of $10^{-22}$. Assuming $u_s=2u_{surf}(+)+2u_{surf}(b)$, we should have $u_s=0$ following Eq. (\ref{eq:usurf-+}) and (\ref{eq:usurf-b}). The surface term in Tab. \ref{Tab:energy-+b+b} $|u_s| < 10^{-17}$ is nearly zero. For the corner term, assuming $u_c=4u_{corn}(+a)$, we conjecture that
\begin{equation}
u_{corn}(+b)=\frac{5\sqrt{2}}{8\pi}
\end{equation}
This is valid in the accuracy of $10^{-15}$ (see Tab.\ref{Tab:energy-+b+b})

\begin{table}
 \caption{ The fitted parameters for the internal energy density with BCs ($a0a0$). It has $\delta_{max}<10^{-29}$.}
\begin{tabular}{crr}

\hline
            &  $u_i$~~~~~~~~~~~~~~~~~~~~~~~~~~~~~      &  $\Delta u_i$ ~~~~\\
\hline
$u_0 $  &   $-$0.141421356237309504880168851193D+01&     0.8D$-$25 \\
$u_s $  &    0.127323954473516268614957785079D+01&     0.5D$-$21 \\
$u_1 $  &    0.121362669205892989049105448341D+00&     0.5D$-$20 \\
$u_c $  &    0.636619772367581340801507421716D+00&     0.7D$-$18 \\
$u_2 $  &    0.575980763410885278095996550463D+00&     0.4D$-$17 \\
$u_3 $  &    0.506763654594585805239275861483D+00&     0.3D$-$15 \\
$u_4 $  &   $-$0.459595022346897357884333871040D+00&     0.5D$-$13 \\
$u_5 $  &   $-$0.741950809610363725073693898886D+00&     0.8D$-$11 \\
$u_6 $  &   $-$0.839344511700144627100201306218D-02&     0.1D$-$08 \\
$u_7 $  &   $-$0.350019466440250214188463383846D+01&     0.1D$-$06 \\
$u_8 $  &    0.663027485274331392787685849194D+00&     0.1D$-$04 \\
$u_9 $  &   $-$0.250237843579355326620032363485D+02&     0.8D$-$03 \\
$u_{10}$  &   $-$0.223448993693708657816392285520D+02&     0.4D$-$01 \\
$u_{11}$  &   $-$0.435152923522988391018566162983D+03&     0.2D+01 \\
$u_{12}$  &   $-$0.192430230145581047691910486037D+03&     0.5D+02 \\
$u_{13}$  &   $-$0.285863055289567306004616541995D+05&     0.8D+03 \\
$u_{14}$  &    0.223092860557318122922870888567D+06&     0.9D+04 \\
$u_{15}$  &   $-$0.253063054430581768311527405196D+07&     0.5D+05 \\

\hline
\end{tabular}
\label{Tab:energy-a0a0}
\end{table}

Tab. \ref{Tab:energy-a0a0} is for the BCs $(a0a0)$. The fitted bulk term $u_0$  agrees with the exact result $u_0=\sqrt{2}$ in the accuracy of $10^{-25}$. Assuming $u_s=2u_{surf}(0)+2u_{surf}(a)$, we should have $u_s=4/\pi$ following Eq. (\ref{eq:usurf-a}) and (\ref{eq:usurf-0}). For the corner term, assuming $u_c=4u_{corn}(a0)$, we conjecture that
\begin{equation}
u_{corn}(a0)=\frac{1}{2\pi}
\end{equation}
This is valid in the accuracy of $10^{-18}$ (see Tab.\ref{Tab:energy-a0a0})

\begin{table}
 \caption{ The fitted parameters for the internal energy density with BCs ($b0b0$). It has $\delta_{max}<10^{-25}$. }
\begin{tabular}{crr}

\hline
            &  $u_i$~~~~~~~~~~~~~~~~~~~~~~~~~~~~~      &  $\Delta u_i$ ~~~~\\
\hline
$u_0 $  &   $-$0.141421356237309504880132870812D+01&     0.8D$-$21 \\
$u_s $  &    0.127323954473516268360487483992D+01&     0.5D$-$17 \\
$u_1 $  &    0.861723523892906483642885887830D-01&     0.4D$-$16 \\
$u_c $  &    0.411540693328300930568165597271D+00&     0.7D$-$14 \\
$u_2 $  &   $-$0.455240659177989954019392349433D+00&     0.4D$-$13 \\
$u_3 $  &    0.137937078637275746526096487554D+01&     0.3D$-$11 \\
$u_4 $  &   $-$0.209914617819527310903232708888D+01&     0.5D$-$09 \\
$u_5 $  &   $-$0.689169258981385034763046369127D+01&     0.8D$-$07 \\
$u_6 $  &   $-$0.129241734582760720996744385809D+02&     0.1D$-$04 \\
$u_7 $  &   $-$0.531077800877101688558806106130D+02&     0.1D$-$02 \\
$u_8 $  &   $-$0.161761716277689928523984437306D+03&     0.1D+00 \\
$u_9 $  &   $-$0.862517747773288349539320678429D+03&     0.8D+01 \\
$u_{10}$  &   $-$0.299280392861972456308739526517D+04&     0.4D+03 \\
$u_{11}$  &   $-$0.616962286932370019659330553037D+05&     0.2D+05 \\
$u_{12}$  &    0.904632672481344443101910870464D+06&     0.5D+06 \\
$u_{13}$  &   $-$0.278347218927750357311487682794D+08&     0.8D+07 \\
$u_{14}$  &    0.357312296468819135474507455962D+09&     0.9D+08 \\
$u_{15}$  &   $-$0.277918465232931366855053206891D+10&     0.5D+09 \\

\hline
\end{tabular}
\label{Tab:energy-b0b0}
\end{table}

Tab. \ref{Tab:energy-b0b0} is for the BCs $(b0b0)$. The fitted bulk term $u_0$  agrees with the exact result $u_0=\sqrt{2}$ in the accuracy of $10^{-21}$. Assuming $u_s=2u_{surf}(0)+2u_{surf}(b)$, we should have $u_s=4/\pi$ following Eq. (\ref{eq:usurf-b}) and (\ref{eq:usurf-0}). For the corner term, assuming $u_c=4u_{corn}(b0)$, we conjecture that
\begin{equation}
u_{corn}(b0)=\frac{4-\sqrt{2}}{8\pi}
\end{equation}
This is valid in the accuracy of $10^{-14}$ (see Tab.\ref{Tab:energy-b0b0})

\begin{table}
 \caption{ The fitted parameters for the internal energy density with BCs ($abab$). It has $\delta_{max}<10^{-26}$.}
\begin{tabular}{crr}

\hline
            &  $u_i$~~~~~~~~~~~~~~~~~~~~~~~~~~~~~      &  $\Delta u_i$ ~~~~\\
\hline
$u_0 $  &   $-$0.141421356237309504880136941429D+01&     0.8D$-$22 \\
$u_s $  &    0.127323954473516268389306332551D+01&     0.5D$-$18 \\
$u_1 $  &    0.861723523892906458635504480884D-01&     0.4D$-$17 \\
$u_c $  &    0.147844149538776645459270272732D+00&     0.7D$-$15 \\
$u_2 $  &   $-$0.719746505435468031604787898128D+00&     0.4D$-$14 \\
$u_3 $  &    0.918823795396608499030357447921D+00&     0.3D$-$12 \\
$u_4 $  &   $-$0.602104474739628494712865252576D+00&     0.5D$-$10 \\
$u_5 $  &   $-$0.593185548579978946841655294667D+01&     0.8D$-$08 \\
$u_6 $  &   $-$0.858376410731878611197966862115D+01&     0.1D$-$05 \\
$u_7 $  &   $-$0.434303527417545744014575647528D+02&     0.1D$-$03 \\
$u_8 $  &   $-$0.130082936660134750887129009656D+03&     0.1D$-$01 \\
$u_9 $  &   $-$0.766892721889005178090740431273D+03&     0.8D+00 \\
$u_{10}$  &   $-$0.258669611192464081711085295718D+04&     0.4D+02 \\
$u_{11}$  &   $-$0.528767912491730915959422973208D+05&     0.2D+04 \\
$u_{12}$  &    0.814250684914679398240826839926D+06&     0.5D+05 \\
$u_{13}$  &   $-$0.245214853763881527781338267171D+08&     0.8D+06 \\
$u_{14}$  &    0.315176776154582420697154062665D+09&     0.9D+07 \\
$u_{15}$  &   $-$0.243974488313371951495364255957D+10&     0.5D+08 \\

\hline
\end{tabular}
\label{Tab:energy-abab}
\end{table}

Tab. \ref{Tab:energy-abab} is for the BCs $(abab)$. The fitted bulk term $u_0$  agrees with the exact result $u_0=\sqrt{2}$ in the accuracy of $10^{-22}$. Assuming $u_s=2u_{surf}(a)+2u_{surf}(b)$, we should have $u_s=4/\pi$ following Eq. (\ref{eq:usurf-b}) and (\ref{eq:usurf-a}). For the corner term, assuming $u_c=4u_{corn}(ab)$, we conjecture that
\begin{equation}
u_{corn}(ab)=\frac{8-5\sqrt{2}}{8\pi}
\end{equation}
This is valid in the accuracy of $10^{-15}$ (see Tab.\ref{Tab:energy-abab}).

It is worth to mention an interesting case, in which the BCs is $(++00)$. There is no finite size correction in the internal energy.
The internal energy density equals to $-\sqrt{2}$ in the accuracy of $10^{-28}$ for $30<N<1000$ in the numerical results. According to the discussion above, the edge and corner terms are expected to be canceled. The surprising thing is that the corrections at every order are canceled. This may be an exact result. This is similar to the cylinder with Brascamp-Kunz BCS \cite{izmailian2002b}, the critical internal energy density is exactly $-\sqrt{2}$.

\section{The critical specific heat}

\begin{table}
 \caption{ The fitted parameters for the specific heat on a square with BCs ($++++$). It has $\delta_{max}<10^{-29}$.}
\begin{tabular}{crr}

\hline
            &  $c_i$~~~~~~~~~~~~~~~~~~~~~~~~~~~~~      &  $\Delta c_i$ ~~~~\\
\hline
$A_0 $  &    0.254647908947032537230214283412D+01&     0.3D$-$23 \\
$D_0 $  &   $-$0.293909347672490423436325725581D+01&     0.3D$-$22 \\
$c_s $  &    0.689359081125486288373149696009D+01&     0.2D$-$19 \\
$c_1 $  &   $-$0.327342738266179158816794038135D+01&     0.1D$-$18 \\
$c_c $  &    0.562035126651970020925659133942D+01&     0.1D$-$16 \\
$c_2 $  &    0.575090362348364281135590572756D+01&     0.6D$-$16 \\
$c_3 $  &    0.298083919674543867389743708012D+01&     0.3D$-$14 \\
$c_4 $  &   $-$0.908487924214544439227266983743D+00&     0.3D$-$12 \\
$c_5 $  &   $-$0.694091729441007061467482414892D+00&     0.4D$-$10 \\
$c_6 $  &    0.475860363194823053390829772522D+01&     0.4D$-$08 \\
$c_7 $  &   $-$0.173606196612742179491953041821D+02&     0.4D$-$06 \\
$c_8 $  &    0.577151477538787992838523598397D+02&     0.3D$-$04 \\
$c_9 $  &   $-$0.195728910039664167205691709740D+03&     0.2D$-$02 \\
$c_{10}$  &    0.714196276906833920017608104383D+03&     0.7D$-$01 \\
$c_{11}$  &   $-$0.289741609822175702189449817833D+04&     0.2D+01 \\
$c_{12}$  &    0.129887873829990655384164670328D+05&     0.5D+02 \\
$c_{13}$  &   $-$0.604027923454997979109739684410D+05&     0.8D+03 \\
$c_{14}$  &    0.244714542486248535036743842440D+06&     0.8D+04 \\
$c_{15}$  &   $-$0.595098651994059987957898043538D+06&     0.4D+05 \\

\hline
\end{tabular}
\label{Tab:specific-++++}
\end{table}

The  critical specific heat can be expanded into
\begin{equation}
c=A_0 \ln N+D_0+c_{\rm surf}\frac{\ln N}{N}
+c_{\rm corn}\frac{\ln S }{S} +\sum_{k=1}^{\infty}\frac{c_k}{N^k}.
\end{equation}
where the surface term $c_s=\sum_{i=1}^{4}c_{\rm surf}(\alpha_i)$ is the sum of the four edges' contribution, the corner term $c_c=\sum_{i=1}^4c_{\rm corn}(\alpha_i\beta_i)$ is the sum of the four corners' contribution.

The  exact value of $A_0$ is known from Onsager's solution \cite{onsager}
\begin{equation}
A_0 =\frac{8}{\pi}=2.546479089470325372302140\cdots
\end{equation}
This is the benchmark of our fittings.

Tab. \ref{Tab:specific-++++} is for the BCs $(++++)$. The fitted bulk term $A_0$  agrees with the exact result in the accuracy of $10^{-23}$. Assuming $c_s=4c_{surf}(+)$, we conjecture that
\begin{equation}
c_{surf}(+)=\frac{4+\sqrt{2}}{\pi}.
\label{eq:csurf-+}
\end{equation}
The surface term in Tab. \ref{Tab:specific-++++} agrees with this conjecture in the accuracy of $10^{-19}$  . For the corner term, assuming $c_c=4c_{corn}(++)$, we conjecture that
\begin{equation}
c_{corn}(++)=\frac{3+\sqrt{2}}{\pi}.
\end{equation}
This is valid in the accuracy of $10^{-16}$ (see Tab.\ref{Tab:specific-++++})

\begin{table}
 \caption{ The fitted parameters for the specific heat on a square with fixed BCs ($aaaa$). It has $\delta_{max}<10^{-28}$.}
\begin{tabular}{crr}

\hline
            &  $c_i$~~~~~~~~~~~~~~~~~~~~~~~~~~~~~      &  $\Delta c_i$ ~~~~\\
\hline
$A_0 $  &    0.254647908947032537230206978923D+01&     0.9D$-$22 \\
$D_0 $  &   $-$0.293909347672490423436250434308D+01&     0.9D$-$21 \\
$c_s $  &    0.329232554662643860499790202488D+01&     0.5D$-$18 \\
$c_1 $  &   $-$0.104684259478955479620773150699D+01&     0.4D$-$17 \\
$c_c $  &    0.201908600189127553855917230430D+01&     0.4D$-$15 \\
$c_2 $  &    0.301111569698128319085688354454D+01&     0.2D$-$14 \\
$c_3 $  &    0.183531726641867451632631519000D+01&     0.1D$-$12 \\
$c_4 $  &   $-$0.271303925099154101453020914221D+00&     0.1D$-$10 \\
$c_5 $  &   $-$0.287411051280984622313740815566D+00&     0.2D$-$08 \\
$c_6 $  &    0.121105203952828033760939117636D+01&     0.2D$-$06 \\
$c_7 $  &   $-$0.799522018121570104184779326334D+01&     0.2D$-$04 \\
$c_8 $  &   $-$0.299648518954361642014171307720D+02&     0.2D$-$02 \\
$c_9 $  &   $-$0.917236609518639320482804562899D+02&     0.1D+00 \\
$c_{10}$  &   $-$0.911065004850311165635153077970D+03&     0.6D+01 \\
$c_{11}$  &   $-$0.324283549568478522491160062243D+04&     0.2D+03 \\
$c_{12}$  &   $-$0.478722445736335054232819689967D+04&     0.6D+04 \\
$c_{13}$  &   $-$0.709727204995004199554925888622D+06&     0.1D+06 \\
$c_{14}$  &    0.739208835251761678649825640593D+07&     0.1D+07 \\
$c_{15}$  &   $-$0.733379769407458591275859667221D+08&     0.5D+07 \\

\hline
\end{tabular}
\label{Tab:specific-aaaa}
\end{table}

The specific heat for the boundary  condition $(0000)$ has been studied in Ref. \cite{wu3}. According to the definition of the present paper, the surface
term is given by
\begin{equation}
c_{surf}(0)=\frac{3\sqrt{2}}{\pi}
\label{eq:csurf-0}
\end{equation}
and the corner term is given by
\begin{equation}
c_{corn}(00)=\frac{3}{\pi}.
\label{eq:ccorn-0}
\end{equation}

Tab. \ref{Tab:specific-aaaa} is for the BCs  $(aaaa)$. The fitted bulk term $A_0$  agrees with the exact result in the accuracy of $10^{-22}$. Assuming $c_s=4c_{surf}(a)$, we conjecture that
\begin{equation}
c_{surf}(a)=\frac{4-\sqrt{2}}{\pi}.
\label{eq:csurf-a}
\end{equation}
The surface term in Tab. \ref{Tab:specific-aaaa} agrees with this conjecture in the accuracy of $10^{-19}$  . For the corner term, assuming $c_c=4c_{corn}(aa)$, we conjecture that
\begin{equation}
c_{corn}(aa)=\frac{3-\sqrt{2}}{\pi}.
\end{equation}
This is valid in the accuracy of $10^{-16}$ (see Tab.\ref{Tab:specific-aaaa})

\begin{table}
 \caption{ The fitted parameters for the specific heat on a square with fixed BCs ($bbbb$). It has $\delta_{max}<10^{-25}$. }
\begin{tabular}{crr}

\hline
            &  $c_i$~~~~~~~~~~~~~~~~~~~~~~~~~~~~~      &  $\Delta c_i$ ~~~~\\
\hline
$A_0 $  &    0.254647908947032537222628362552D+01&     0.8D$-$19 \\
$D_0 $  &   $-$0.293909347672490423358062420702D+01&     0.8D$-$18 \\
$c_s $  &    0.149169291431222592937112453344D+01&     0.5D$-$15 \\
$c_1 $  &    0.960720968523530075669626709447D+00&     0.4D$-$14 \\
$c_c $  &    0.181000280049560355823967811254D+01&     0.4D$-$12 \\
$c_2 $  &    0.481367310523746713408754687571D+01&     0.2D$-$11 \\
$c_3 $  &    0.123909793524648500484705119479D+01&     0.9D$-$10 \\
$c_4 $  &    0.744642721151599474544853049262D+01&     0.1D$-$07 \\
$c_5 $  &    0.355000045455129584728604259343D+01&     0.2D$-$05 \\
$c_6 $  &    0.998735014426295844300108432970D+01&     0.2D$-$03 \\
$c_7 $  &    0.114487447556004201204470374903D+02&     0.2D$-$01 \\
$c_8 $  &   $-$0.181446304927090746187979805657D+03&     0.2D+01 \\
$c_9 $  &   $-$0.307396103050974932037373508290D+04&     0.1D+03 \\
$c_{10}$  &   $-$0.271970783299738997302606837840D+04&     0.6D+04 \\
$c_{11}$  &   $-$0.100442694505996843617605591754D+07&     0.2D+06 \\
$c_{12}$  &    0.239042798834460583609749732933D+08&     0.6D+07 \\
$c_{13}$  &   $-$0.569423487579979621261925932548D+09&     0.9D+08 \\
$c_{14}$  &    0.735090079314670933702164285297D+10&     0.1D+10 \\
$c_{15}$  &   $-$0.507545172234954561599039280458D+11&     0.5D+10 \\

\hline
\end{tabular}
\label{Tab:specific-bbbb}
\end{table}

Tab. \ref{Tab:specific-bbbb} is for the BCs $(bbbb)$. The fitted bulk term $A_0$  agrees with the exact result in the accuracy of $10^{-19}$. Assuming $c_s=4c_{surf}(b)$, we conjecture that
\begin{equation}
c_{surf}(b)=\frac{4-2\sqrt{2}}{\pi}.
\label{eq:csurf-b}
\end{equation}
The surface term in Tab. \ref{Tab:specific-bbbb} agrees with this conjecture in the accuracy of $10^{-15}$  . For the corner term, assuming $c_c=4c_{corn}(bb)$, we conjecture that
\begin{equation}
c_{corn}(bb)=\frac{17-8\sqrt{2}}{4\pi}.
\end{equation}
This is valid in the accuracy of $10^{-12}$ (see Tab.\ref{Tab:specific-bbbb})

\begin{table}
 \caption{ The fitted parameters for the specific heat on a square with fixed BCs ($+-+-$). It has $\delta_{max}<10^{-29}$.}
\begin{tabular}{crr}
\hline
            &  $c_i$~~~~~~~~~~~~~~~~~~~~~~~~~~~~~      &  $\Delta c_i$ ~~~~\\
\hline
$A_0 $  &    0.254647908947032537236144399742D+01&     0.3D$-$23 \\
$D_0 $  &   $-$0.492727482999274702712092394042D+01&     0.3D$-$22 \\
$c_s $  &    0.689359081125486326825718578894D+01&     0.2D$-$19 \\
$c_1 $  &   $-$0.150452609201860687597073036122D+01&     0.1D$-$18 \\
$c_c $  &    0.562035126651997040563182146608D+01&     0.1D$-$16 \\
$c_2 $  &    0.116220489712764056399214578988D+02&     0.6D$-$16 \\
$c_3 $  &   $-$0.119910298798203650394259147957D+02&     0.3D$-$14 \\
$c_4 $  &    0.200554210944858037782029635894D+02&     0.3D$-$12 \\
$c_5 $  &   $-$0.122901248267278239210199284093D+03&     0.4D$-$10 \\
$c_6 $  &    0.801535418478596493995308197027D+03&     0.4D$-$08 \\
$c_7 $  &   $-$0.625715877020679208326957704801D+04&     0.4D$-$06 \\
$c_8 $  &    0.402700617678530584896555284140D+05&     0.3D$-$04 \\
$c_9 $  &   $-$0.264504692533820697537699316172D+06&     0.2D$-$02 \\
$c_{10}$  &    0.182300386675525493056049729107D+07&     0.7D$-$01 \\
$c_{11}$  &   $-$0.138669232341746208323906454620D+08&     0.2D+01 \\
$c_{12}$  &    0.105903860981512403270780377819D+09&     0.5D+02 \\
$c_{13}$  &   $-$0.747932457741706237920148902789D+09&     0.8D+03 \\
$c_{14}$  &    0.404099776087486211121179900436D+10&     0.8D+04 \\
$c_{15}$  &   $-$0.123613019366722099517279596642D+11&     0.4D+05 \\

\hline
\end{tabular}
\label{Tab:specific-+-+-}
\end{table}

Tab. \ref{Tab:specific-+-+-} is for the BCs $(+-+-)$. The fitted bulk term $A_0$  agrees with the exact result in the accuracy of $10^{-23}$. Assuming $c_s=2c_{surf}(+)+2c_{surf}(-)$, and $c_{surf}(-)=c_{surf}(+)$, we have $c_s=4(4+\sqrt{2})/\pi$ following Eq. (\ref{eq:csurf-+}).
The surface term in Tab. \ref{Tab:specific-+-+-} agrees with this conjecture in the accuracy of $10^{-19}$  . For the corner term, assuming $c_c=4c_{corn}(+-)$, we conjecture that
\begin{equation}
c_{corn}(+-)=\frac{3+\sqrt{2}}{\pi}.
\end{equation}
This is valid in the accuracy of $10^{-16}$ (see Tab.\ref{Tab:specific-+-+-})

\begin{table}
 \caption{ The fitted parameters for the specific heat on a square with fixed BCs ($+0+0$). It has $\delta_{max}<10^{-29}$.}
\begin{tabular}{crr}
\hline
            &  $c_i$~~~~~~~~~~~~~~~~~~~~~~~~~~~~~      &  $\Delta c_i$ ~~~~\\
\hline
$A_0 $  &    0.254647908947032537230212991908D+01&     0.3D$-$23 \\
$D_0 $  &   $-$0.207957263615531854911636408755D+01&     0.3D$-$22 \\
$c_s $  &    0.614774435409874965045620596750D+01&     0.2D$-$19 \\
$c_1 $  &   $-$0.101309734239104595716631093227D+01&     0.1D$-$18 \\
$c_c $  &    0.270094894847131816182160626168D+01&     0.1D$-$16 \\
$c_2 $  &    0.526805790482609384386607630965D+01&     0.6D$-$16 \\
$c_3 $  &   $-$0.249753387885966217510362245280D+01&     0.3D$-$14 \\
$c_4 $  &    0.370691815885660939491372334292D+01&     0.3D$-$12 \\
$c_5 $  &    0.941507246344439813215732887542D+00&     0.4D$-$10 \\
$c_6 $  &   $-$0.310081533788721514302149061280D+01&     0.4D$-$08 \\
$c_7 $  &   $-$0.176102487428439225260161700682D+02&     0.4D$-$06 \\
$c_8 $  &    0.658977981685206056374208493495D+02&     0.3D$-$04 \\
$c_9 $  &    0.348414635156862573811609082290D+02&     0.2D$-$02 \\
$c_{10}$  &   $-$0.108636519768285864753238035323D+04&     0.7D$-$01 \\
$c_{11}$  &    0.661108338127640618317264635223D+04&     0.2D+01 \\
$c_{12}$  &   $-$0.352033143363045929771067881519D+05&     0.5D+02 \\
$c_{13}$  &    0.198754824807194272686385157738D+06&     0.8D+03 \\
$c_{14}$  &   $-$0.938196322052719094417967872397D+06&     0.8D+04 \\
$c_{15}$  &    0.264028811854196437647623304356D+07&     0.4D+05 \\

\hline
\end{tabular}
\label{Tab:specific-+0+0}
\end{table}

Tab. \ref{Tab:specific-+0+0} is for the BCs $(+0+0)$. The fitted bulk term $A_0$  agrees with the exact result in the accuracy of $10^{-23}$. Assuming $c_s=2c_{surf}(+)+2c_{surf}(0)$,  we should have $c_s=8(1+\sqrt{2})/\pi$ following Eq. (\ref{eq:csurf-+}) and (\ref{eq:csurf-0}).
The surface term in Tab. \ref{Tab:specific-+0+0} agrees with this conjecture in the accuracy of $10^{-19}$  . For the corner term, assuming $c_c=4c_{corn}(+0)$, we conjecture that
\begin{equation}
c_{corn}(+0)=\frac{3\sqrt{2}}{2\pi}.
\end{equation}
This is valid in the accuracy of $10^{-16}$ (see Tab.\ref{Tab:specific-+0+0}).

\begin{table}
 \caption{ The fitted parameters for the specific heat on a square with fixed BCs ($+a+a$). It has $\delta_{max}<10^{-29}$. }
\begin{tabular}{crr}
\hline
            &  $c_i$~~~~~~~~~~~~~~~~~~~~~~~~~~~~~      &  $\Delta c_i$ ~~~~\\
\hline
$A_0 $  &    0.254647908947032537230210574312D+01&     0.9D$-$23 \\
$D_0 $  &   $-$0.207957263615531854911611392444D+01&     0.9D$-$22 \\
$c_s $  &    0.509295817894065074436098580238D+01&     0.5D$-$19 \\
$c_1 $  &    0.189228665387521176913999928370D+00&     0.4D$-$18 \\
$c_c $  &    0.127323954473516249894242321920D+01&     0.4D$-$16 \\
$c_2 $  &    0.314495167832656126270550022997D+01&     0.2D$-$15 \\
$c_3 $  &   $-$0.361306019538510057178978293970D+01&     0.1D$-$13 \\
$c_4 $  &    0.321987660381074099218342524838D+01&     0.1D$-$11 \\
$c_5 $  &    0.100947036600403325684642136691D+02&     0.2D$-$09 \\
$c_6 $  &   $-$0.125252866676884385924280980631D+02&     0.2D$-$07 \\
$c_7 $  &   $-$0.516423179150607941402913785860D+02&     0.2D$-$05 \\
$c_8 $  &    0.578774177487123299474878273119D+02&     0.2D$-$03 \\
$c_9 $  &    0.203470309210880925538109936354D+03&     0.1D$-$01 \\
$c_{10}$  &   $-$0.256998525530611434748477701389D+04&     0.6D+00 \\
$c_{11}$  &    0.680299531994107434185852305362D+04&     0.2D+02 \\
$c_{12}$  &   $-$0.546180237741634618076393149966D+05&     0.6D+03 \\
$c_{13}$  &   $-$0.742271643234979729556882150352D+05&     0.1D+05 \\
$c_{14}$  &    0.201644788170527883119266622262D+07&     0.1D+06 \\
$c_{15}$  &   $-$0.319518039737334413809390744112D+08&     0.5D+06 \\

\hline
\end{tabular}
\label{Tab:specific-+a+a}
\end{table}

Tab. \ref{Tab:specific-+a+a} is for the BCS $(+a+a)$. The fitted bulk term $A_0$  agrees with the exact result in the accuracy of $10^{-23}$. Assuming $c_s=2c_{surf}(+)+2c_{surf}(a)$,  we should have $c_s=16/\pi$ following Eq. (\ref{eq:csurf-+}) and (\ref{eq:csurf-a}).
The surface term in Tab. \ref{Tab:specific-+a+a} agrees with this conjecture in the accuracy of $10^{-19}$  . For the corner term, assuming $c_c=4c_{corn}(+a)$, we conjecture that
\begin{equation}
c_{corn}(+a)=\frac{1}{\pi}.
\end{equation}
This is valid in the accuracy of $10^{-16}$ (see Tab.\ref{Tab:specific-+a+a}).

\begin{table}
 \caption{ The fitted parameters for the specific heat on a square with fixed BCs ($+b+b$). It has $\delta_{max}<10^{-26}$. }
\begin{tabular}{crr}
\hline
            &  $c_i$~~~~~~~~~~~~~~~~~~~~~~~~~~~~~      &  $\Delta c_i$ ~~~~\\
\hline
$A_0 $  &    0.254647908947032537229448304623D+01&     0.8D$-$20 \\
$D_0 $  &   $-$0.207957263615531854903747176070D+01&     0.8D$-$19 \\
$c_s $  &    0.419264186278354462084675833691D+01&     0.5D$-$16 \\
$c_1 $  &    0.113029158169382278736450277193D+01&     0.4D$-$15 \\
$c_c $  &   $-$0.263696543789566466441289778314D+00&     0.4D$-$13 \\
$c_2 $  &    0.731585847405046478725572017121D+00&     0.2D$-$12 \\
$c_3 $  &   $-$0.456887964199219970935770665880D+01&     0.9D$-$11 \\
$c_4 $  &   $-$0.348673125857667659825142520665D+01&     0.1D$-$08 \\
$c_5 $  &    0.276240284580850486393029244801D+02&     0.2D$-$06 \\
$c_6 $  &    0.151916665950816568173252489517D+02&     0.2D$-$04 \\
$c_7 $  &   $-$0.872021250638779320602164906733D+02&     0.2D$-$02 \\
$c_8 $  &   $-$0.279715966875344244925819167108D+03&     0.2D+00 \\
$c_9 $  &   $-$0.416404995498745325644835577058D+03&     0.1D+02 \\
$c_{10}$  &   $-$0.637254699732866821468327034824D+04&     0.6D+03 \\
$c_{11}$  &   $-$0.112855773947155150769207171356D+06&     0.2D+05 \\
$c_{12}$  &    0.217333921165144399861602110101D+07&     0.6D+06 \\
$c_{13}$  &   $-$0.588795974877262539291820110828D+08&     0.9D+07 \\
$c_{14}$  &    0.744195754263875360561031278706D+09&     0.1D+09 \\
$c_{15}$  &   $-$0.543196548585847101320613150355D+10&     0.5D+09 \\

\hline
\end{tabular}
\label{Tab:specific-+b+b}
\end{table}

Tab. \ref{Tab:specific-+b+b} is for the BCs $(+b+b)$. The fitted bulk term $A_0$  agrees with the exact result in the accuracy of $10^{-20}$. Assuming $c_s=2c_{surf}(+)+2c_{surf}(b)$,  we should have $c_s=2(8-\sqrt{2})/\pi$ following Eq. (\ref{eq:csurf-+}) and (\ref{eq:csurf-b}).
The surface term in Tab. \ref{Tab:specific-+b+b} agrees with this conjecture in the accuracy of $10^{-16}$  . For the corner term, assuming $c_c=4c_{corn}(+b)$, we conjecture that
\begin{equation}
c_{corn}(+b)=\frac{1-\sqrt{2}}{2\pi}.
\end{equation}
This is valid in the accuracy of $10^{-13}$ (see Tab.\ref{Tab:specific-+b+b}).

\begin{table}
 \caption{ The fitted parameters for the specific heat on a square with fixed BCs ($a0a0$). It has $\delta_{max}<10^{-29}$.  }
\begin{tabular}{crr}
\hline
            &  $c_i$~~~~~~~~~~~~~~~~~~~~~~~~~~~~~      &  $\Delta c_i$ ~~~~\\
\hline
$A_0 $  &    0.254647908947032537230210471955D+01&     0.9D$-$23 \\
$D_0 $  &   $-$0.293909347672490423436286472184D+01&     0.9D$-$22 \\
$c_s $  &    0.434711172178453751116404076485D+01&     0.5D$-$19 \\
$c_1 $  &   $-$0.232264832397915819750369604603D+01&     0.4D$-$18 \\
$c_c $  &    0.270094894847131801889630886119D+01&     0.4D$-$16 \\
$c_2 $  &    0.391383478029488326699906509162D+01&     0.2D$-$15 \\
$c_3 $  &    0.164147182373922418309343319991D+01&     0.1D$-$13 \\
$c_4 $  &    0.378615173373695852119617699393D+00&     0.1D$-$11 \\
$c_5 $  &   $-$0.796287269422202748959087573965D+00&     0.2D$-$09 \\
$c_6 $  &    0.410840448192540613778228578418D+01&     0.2D$-$07 \\
$c_7 $  &   $-$0.102560916958324198606430729837D+02&     0.2D$-$05 \\
$c_8 $  &    0.669107880644565672652132047658D+01&     0.2D$-$03 \\
$c_9 $  &   $-$0.103081357008300759809552994588D+03&     0.1D$-$01 \\
$c_{10}$  &   $-$0.180312137400539556044710683501D+03&     0.6D+00 \\
$c_{11}$  &   $-$0.252436551750684903817000851523D+04&     0.2D+02 \\
$c_{12}$  &    0.336383493897803788305010742555D+04&     0.6D+03 \\
$c_{13}$  &   $-$0.370345611670357382424544011290D+06&     0.1D+05 \\
$c_{14}$  &    0.372759208799472356330611043872D+07&     0.1D+06 \\
$c_{15}$  &   $-$0.358676104518076416664723664113D+08&     0.5D+06 \\

\hline
\end{tabular}
\label{Tab:specific-a0a0}
\end{table}

Tab. \ref{Tab:specific-a0a0} is for the BCs $(a0a0)$. The fitted bulk term $A_0$  agrees with the exact result in the accuracy of $10^{-23}$. Assuming $c_s=2c_{surf}(a)+2c_{surf}(0)$,  we should have $c_s=2(4+2\sqrt{2})/\pi$ following Eq. (\ref{eq:csurf-0}) and (\ref{eq:csurf-a}).
The surface term in Tab. \ref{Tab:specific-a0a0} agrees with this conjecture in the accuracy of $10^{-19}$  . For the corner term, assuming $c_c=4c_{corn}(a0)$, we conjecture that
\begin{equation}
c_{corn}(a0)=\frac{3\sqrt{2}}{2\pi}.
\end{equation}
This is valid in the accuracy of $10^{-16}$ (see Tab.\ref{Tab:specific-a0a0}).

\begin{table}
 \caption{ The fitted parameters for the specific heat on a square with fixed BCs ($b0b0$). It has $\delta_{max}<10^{-25}$.}
\begin{tabular}{crr}
\hline
            &  $c_i$~~~~~~~~~~~~~~~~~~~~~~~~~~~~~      &  $\Delta c_i$ ~~~~\\
\hline
$A_0 $  &    0.254647908947032537222059330773D+01&     0.8D$-$19 \\
$D_0 $  &   $-$0.293909347672490423352191307400D+01&     0.8D$-$18 \\
$c_s $  &    0.344679540562743086452624563398D+01&     0.5D$-$15 \\
$c_1 $  &   $-$0.131886654232261338966254177310D+01&     0.4D$-$14 \\
$c_c $  &    0.206432917610329220276757979748D+01&     0.4D$-$12 \\
$c_2 $  &    0.201742883481703877151623295723D+01&     0.2D$-$11 \\
$c_3 $  &    0.365761842732756826681510852319D+01&     0.9D$-$10 \\
$c_4 $  &    0.987379166874021910384766811308D+01&     0.1D$-$07 \\
$c_5 $  &    0.151694669818126575972225884081D+02&     0.2D$-$05 \\
$c_6 $  &    0.619621461382827653711790166066D+02&     0.2D$-$03 \\
$c_7 $  &    0.762545562731988962727305995223D+02&     0.2D$-$01 \\
$c_8 $  &    0.736115511580950994556990313930D+02&     0.2D+01 \\
$c_9 $  &   $-$0.281272473057038633090930560960D+04&     0.1D+03 \\
$c_{10}$  &    0.217504631303605053453606158282D+04&     0.6D+04 \\
$c_{11}$  &   $-$0.102997599664895788742613206295D+07&     0.2D+06 \\
$c_{12}$  &    0.259790649187562660383344379077D+08&     0.6D+07 \\
$c_{13}$  &   $-$0.611497034869214374808188043862D+09&     0.9D+08 \\
$c_{14}$  &    0.792887411091852955679970893926D+10&     0.1D+10 \\
$c_{15}$  &   $-$0.544433837059006626064505157214D+11&     0.5D+10 \\

\hline
\end{tabular}
\label{Tab:specific-b0b0}
\end{table}

Tab. \ref{Tab:specific-b0b0} is for the BCs $(b0b0)$. The fitted bulk term $A_0$  agrees with the exact result in the accuracy of $10^{-19}$. Assuming $c_s=2c_{surf}(b)+2c_{surf}(0)$,  we should have $c_s=2(4+\sqrt{2})/\pi$ following Eq. (\ref{eq:csurf-0}) and (\ref{eq:csurf-b}).
The surface term in Tab. \ref{Tab:specific-b0b0} agrees with this conjecture in the accuracy of $10^{-15}$  . For the corner term, assuming $c_c=4c_{corn}(b0)$, we conjecture that
\begin{equation}
c_{corn}(b0)=\frac{3\sqrt{2}-1}{2\pi}.
\end{equation}
This is valid in the accuracy of $10^{-12}$ (see Tab.\ref{Tab:specific-b0b0}).

\begin{table}
 \caption{ The fitted parameters for the specific heat on a square with fixed BCs ($abab$). It has $\delta_{max}<10^{-25}$. }
\begin{tabular}{crr}
\hline
            &  $c_i$~~~~~~~~~~~~~~~~~~~~~~~~~~~~~      &  $\Delta c_i$ ~~~~\\
\hline
$A_0 $  &    0.254647908947032537222898804198D+01&     0.8D$-$19 \\
$D_0 $  &   $-$0.293909347672490423360853619665D+01&     0.8D$-$18 \\
$c_s $  &    0.239200923046933201814891396573D+01&     0.5D$-$15 \\
$c_1 $  &   $-$0.430608131330104476792216621391D-01&     0.4D$-$14 \\
$c_c $  &    0.175538945810135247913272226466D+01&     0.4D$-$12 \\
$c_2 $  &    0.197209062011897145411975466469D+01&     0.2D$-$11 \\
$c_3 $  &    0.239822986419313125728026016737D+01&     0.9D$-$10 \\
$c_4 $  &    0.413095167457725306820037301109D+01&     0.1D$-$07 \\
$c_5 $  &    0.135159431639614137459156677669D+02&     0.2D$-$05 \\
$c_6 $  &    0.212985835221577203062026396011D+02&     0.2D$-$03 \\
$c_7 $  &    0.705871113070814927566641932588D+02&     0.2D$-$01 \\
$c_8 $  &   $-$0.682424256106442478290561491838D+02&     0.2D+01 \\
$c_9 $  &   $-$0.192703474411133747023984238303D+04&     0.1D+03 \\
$c_{10}$  &   $-$0.123477145931160971986141262645D+03&     0.6D+04 \\
$c_{11}$  &   $-$0.915776209505705740670104881236D+06&     0.2D+06 \\
$c_{12}$  &    0.231451023489432476221952537471D+08&     0.6D+07 \\
$c_{13}$  &   $-$0.544781936847292123748534543081D+09&     0.9D+08 \\
$c_{14}$  &    0.705467402812520922460014459115D+10&     0.1D+10 \\
$c_{15}$  &   $-$0.483476519699037323679889926889D+11&     0.5D+10 \\
\hline
\end{tabular}
\label{Tab:specific-abab}
\end{table}

Tab. \ref{Tab:specific-abab} is for the BCs $(abab)$. The fitted bulk term $A_0$  agrees with the exact result in the accuracy of $10^{-19}$. Assuming $c_s=2c_{surf}(b)+2c_{surf}(a)$,  we should have $c_s=2(8-3\sqrt{2})/\pi$ following Eq. (\ref{eq:csurf-a}) and (\ref{eq:csurf-b}).
The surface term in Tab. \ref{Tab:specific-abab} agrees with this conjecture in the accuracy of $10^{-15}$  . For the corner term, assuming $c_c=4c_{corn}(ab)$, we conjecture that
\begin{equation}
c_{corn}(ab)=\frac{7-3\sqrt{2}}{2\pi}.
\end{equation}
This is valid in the accuracy of $10^{-12}$ (see Tab.\ref{Tab:specific-abab}).

For the constant term $D_0$, it has
\begin{equation}
D_0 \approx  2.9390934767249042
\end{equation} for the BCs $(++++)$, $(aaaa)$, $(bbbb)$, $(a0a0)$, $(b0b0)$, $(abab)$;
\begin{equation}
D_0\approx  2.079572636155318549
\end{equation}
for $(+0+0)$, $(+a+a)$, $(+b+b)$ and
\begin{equation}
D_0 \approx  4.927274829992747027120
\end{equation} for $ (+-+-) $ respectively. This indicates that boundary ``$0,a,b$" belong the same type in this term.

\section{Discussion and acknowledgment}
We have studied the critical free energy, internal energy and specific heat on square lattices with different fixed boundaries. Due to the extraordinarily high accuracy, we have conjectured all the exact edge and corner logarithmic corrections. These results indicate that for the Ising model with fixed boundary conditions may have analytical solutions.

To be convenient, we only study the square shape because our emphasis is the edge and corner logarithmic corrections. To study the geometry's effect, one should to study different aspect ratio (of the horizontal size and the vertical size). In the previous work \cite{wu,wu2,wu3}, the logarithmic corrections are geometry independent for the Ising model with free BCs. We have not studied the geometry's effects for the BCs in this paper. This is waiting for the future research.

This work is supported by the National Science Foundation of China (NSFC)
under Grant No. 11175018. The author N.I. was also supported by FP7 EU IRSES Project No. 295302 Statistical Physics in
Diverse Realizations and by a Marie Curie International Incoming Fellowship within
the 7th European Community Framework Programme.

\end{document}